\documentclass[10pt,journal]{IEEEtran}

\usepackage[caption=false,font=normalsize,labelfont=sf,textfont=sf]{subfig}
\usepackage{textcomp}
\usepackage{url}
\usepackage{verbatim}
\usepackage{enumitem}

\usepackage{times}
\usepackage{graphicx}
\usepackage[nocompress]{cite}
\IfFileExists{citesort.sty}{\usepackage{citesort}}{}
\usepackage{xcolor}
\usepackage{psfrag}
\usepackage{amssymb}
\usepackage{multirow}
\usepackage{stfloats}
\usepackage{epsfig}
\usepackage{pifont}
\usepackage{amsmath}

\usepackage{cases}
\usepackage{array}
\usepackage{multicol}
\usepackage{multirow}
\usepackage[T1]{fontenc}
\usepackage{comment}
\usepackage{makecell}
\usepackage{amsthm}
\usepackage{indentfirst}
\usepackage{setspace}
\usepackage{bm}
\usepackage{float}
\usepackage{cases}
\usepackage[ruled, lined, linesnumbered, commentsnumbered, longend]{algorithm2e}
\SetKwInOut{KwIn}{Input}
\SetKwInOut{KwOut}{Output}
\SetKwProg{Init}{[Initialization]}{}{end}
\SetKwProg{QT}{[Applying the ELA method]}{}{end}
\SetKwProg{DT}{[Outer Layer Iteration]}{}{end}
\SetKwProg{LT}{[Inner Layer Iteration]}{}{end}
\SetKwProg{QB}{[Iteratively implementing the optimization at each RF]}{}{end}
\graphicspath{{figures/}}
\theoremstyle{remark}
\newtheorem{remark}{Remark}
\makeatletter
\renewcommand{\maketag@@@}[1]{\hbox{\m@th\normalsize\normalfont#1}}
\makeatother
\renewenvironment{thebibliography}[1]{
	\begin{oldthebibliography}{#1}
	}
	{
	\end{oldthebibliography}
}

\makeatletter
\def\ifundefined{\@ifundefined}
\makeatother

\newcommand{\tr}{\text{tr}}
\newcommand{\diag}{\text{diag}}
\newcommand{\EE}{\mathbb{E}}
\newcommand{\R}{\mathbb{R}}

\makeatletter
\g@addto@macro\normalsize{%
}
\g@addto@macro\small{%
}
\makeatother
\begin{document}

		\title{Queue-Aware Graph Reinforcement Learning for UAV-ISAC-Assisted Maritime Data Collection}

	\author{Bohan Li,~\textit{Member, IEEE}, Min Ye, Haochen Liu, Yongkang Gong,~\textit{Member, IEEE}, Ning Gao,~\textit{Member, IEEE}, Jie Nie,~\textit{Member, IEEE}, Pei Xiao,~\textit{Senior Member, IEEE}, Xiuzhen Cheng,~\textit{Fellow, IEEE}  %
		%\thanks{This work was supported in part by the National Key Research and Development Program of China under Grant 2024YFC3109100, in part by the National Natural Science Foundation of China under Grant 62401530, and in part by the Natural Science Foundation of Shandong Province under Grant 2026HWYQ-027 and ZR2024QF076. ({\em Corresponding author: Min Ye.})}
		\thanks{B. Li, M. Ye and J. Nie are with the Faculty of Information Science and Engineering, the Engineering Research Center of Advanced Marine Physical Instruments and Equipment (Ministry of Education), and Qingdao Key Laboratory of Optics and Optoelectronics, Ocean University of China, Qingdao 266100, China (emails: \{bohan.li, yemin, niejie\}@ouc.edu.cn). }%
		\thanks{H. Liu is with the School of Electronics and Information, Northwestern Polytechnical University, Xi'an 710019, China (email: haochenliu@nwpu.edu.cn).}
		\thanks{Y. Gong and X. Cheng are with the School
			of Computer Science and Technology, Shandong University, Qingdao 266100, China
			(email: \{gokawa, xzcheng\}@sdu.edu.cn).}
		\thanks{N. Gao is with the School of Cyber Science and Engineering, Southeast University, Nanjing 210096, China (email: ninggao@seu.edu.cn).}
		\thanks{P. Xiao is with 5GIC \& 6GIC, University of Surrey, Guildford GU2 7XH, U.K. (email: p.xiao@surrey.ac.uk).} %
	}

\IEEEtitleabstractindextext{%
\begin{abstract}
This paper studies high-altitude platform (HAP)-assisted sparse cooperative integrated sensing and communication (ISAC) for UAV-enabled maritime data collection. The model couples a spatially correlated sea-patch field with drifting-buoy dynamics, echo sensing dependent on radar cross section (RCS) and clutter, fused posterior Cram\'er--Rao bounds (PCRBs), and buffered data service. We formulate a long-horizon queue-weighted objective with association, capacity, mobility, and sensing-power constraints enforced by construction, while {PCRB and rate requirements enter as normalized costs}. To solve the resulting mixed discrete--continuous problem, we propose structured feasible-preserving graph multi-agent reinforcement learning (SFPG-MARL). It combines a heterogeneous graph encoder, a prefix-conditioned sequential $b$-matching policy for feasible variable-cardinality association, parameter-shared UAV actors, and a causal centralized critic. Simulations show that SFPG-MARL achieves the best overall sensing--service balance. Under the default setting, its penalized utility is approximately $30\%$ higher than the strongest learning-based alternative and $75\%$ higher than the strongest deterministic scheduler, while the weighted PCRB penalty remains nearly negligible. Ablations identify complementary benefits from relational aggregation, adaptive variable-cardinality association, and proximal policy optimization (PPO) updates. Without retraining, the policy retains its lead when the network size doubles and the mission horizon increases fivefold. It also remains robust across increasingly severe wave-height and sustained-current conditions.
\end{abstract}
}

\maketitle
\IEEEdisplaynontitleabstractindextext
\IEEEpeerreviewmaketitle

\section{Introduction}
Ocean monitoring networks underpin environmental observation and maritime economy by using distributed surface buoys and sensors to acquire oceanographic data over vast, infrastructure-sparse sea areas~\cite{Xia2020MaritimeIoT}. The lack of reliable offshore backhaul, together with constrained link and energy budgets at these terminals, makes timely data delivery difficult and motivates aerial relays that can approach them on demand~\cite{Nomikos2023MaritimeSurvey}. Unmanned aerial vehicles (UAVs) are especially attractive because their controllable mobility enables strong line-of-sight uplinks to dispersed buoys and efficient collection of buffered observations. Meanwhile, integrated sensing and communication (ISAC), which reuses common spectrum and hardware resources for both functions, has emerged as a key sixth-generation (6G) capability~\cite{Liu2022ISACJSAC}. Equipping maritime UAVs with ISAC enables a single platform to localize buoys and collect their data within a shared resource budget, matching the coupled sensing-and-collection needs of ocean monitoring~\cite{Li2026MaritimeMonitor}.

Realizing maritime UAV-ISAC requires trajectory and resource decisions to be jointly optimized because UAV positions shape both the sensing geometry and uplink channel quality, while sensing-power allocation and sparse UAV--buoy association determine how limited resources are distributed among buoys~\cite{Li2025MaritimeEmergency}. Moreover, dynamic maritime conditions intensify this coupling. Specifically, buoy drift continuously changes the relative geometry, so causal control must rely on predicted buoy states rather than fixed terminal locations. At the same time, sea-state-dependent clutter degrades the reliability of echo-based state estimation~\cite{GregersHansen2012SeaClutter}. {Thus, optimizing only current sensing or communication performance may delay urgent data collection and reduce future service opportunities.} These spatial, resource, and temporal dependencies motivate the long-horizon joint optimization studied in this work.

\subsection{Related Works}\label{subsec:related}
\subsubsection{UAV-Enabled ISAC}
UAV-enabled ISAC has been widely investigated as a way to exploit the controllable mobility and line-of-sight channels of aerial platforms for joint sensing and communication. Representative designs jointly optimize UAV maneuvers and transmit beamforming~\cite{Lyu2023UAVISAC}, maximize communication throughput under periodic sensing constraints~\cite{Meng2023PeriodicISAC}, and jointly plan multi-UAV trajectories and resource allocation while bounding target-localization error~\cite{Pan2024UAVISAC}. Multi-UAV designs further coordinate trajectories and resource allocation to improve cooperative detection and robustness~\cite{wang2026isac}. Across these efforts, however, the data actually buffered at the served terminals is commonly abstracted away. Moreover, most studies consider quasi-static terrestrial users or targets and optimize aggregate sensing- or rate-oriented objectives without persistent terminal queues, leaving the maritime setting and its long-horizon data-clearance requirements largely unexplored.

\subsubsection{Maritime UAV-ISAC and Data Collection}
Recently, a growing number of works bring UAVs into the maritime domain. Energy-efficient trajectory designs have been proposed for UAV-aided maritime data collection~\cite{Zhang2022MaritimeWind}, and ISAC has recently been introduced into maritime monitoring networks~\cite{Li2026MaritimeMonitor}. Related studies optimize shipboard maritime ISAC deployment~\cite{Zhang2024MaritimeISAC}, enhance maritime joint sensing and communication with reconfigurable surfaces~\cite{Cao2024IRSMaritime}, and schedule cooperative UAV data collection to reduce the age of information~\cite{Fu2024MaritimeDRL}. Together, these studies establish complementary roles for UAV data collection and maritime ISAC. Beyond low-altitude UAVs, high-altitude platform (HAP) networks can extend wide-area connectivity over infrastructure-sparse regions~\cite{KarabulutKurt2021HAPS}. This capability motivates using the HAP as a fusion-and-coordination node rather than only as a relay. Nevertheless, a system-level maritime formulation that links HAP-assisted multi-UAV sensing with buffered data collection under time-varying ocean conditions remains underexplored.

\subsubsection{Learning-Based Trajectory and Resource Optimization}

The mixed discrete--continuous decisions, nonconvex coupling, and partially observed dynamics in maritime trajectory and resource optimization make repeated online solution by conventional methods difficult, motivating learning-based control. Deep reinforcement learning (DRL) has been applied to UAV-ISAC trajectory and resource control~\cite{Liu2025UAVISACDRL,Zhang2023JRCMultiUAV}. Broader space--air--ground studies also reveal strong cross-domain coupling among sensing, communication, and computation resources~\cite{Mao2024SAGINISAC}. Graph neural networks (GNNs) effectively exploit the relational structure of wireless networks~\cite{Shen2023GNNWireless}, and their integration with multi-agent reinforcement learning (MARL) supports scalable and topology-aware coordination~\cite{Liu2024GNNCommMARL}. Recent GNN-MARL designs use a multi-relational heterogeneous graph with centralized training and decentralized execution (CTDE) to schedule layered-UAV trajectories and data offloading~\cite{Zhao2026HGMARL}, or a hierarchical graph policy to coordinate association, radio resources, and trajectories in multi-UAV ISAC~\cite{Wang2026HGMRL}. Complementary scalable MARL studies combine attention and parameter sharing for variable-size multi-UAV coverage~\cite{Chen2024TMARL}, or employ a commander--worker hierarchy with centralized-attention critics for access point (AP)--user pairing and bandwidth--power allocation in cell-free ISAC~\cite{Jiang2026HADMARL}. These studies demonstrate the value of relational encoding and centralized coordination. However, dynamic maritime data collection additionally couples capacity-constrained sparse UAV--buoy association with continuous mobility and resource decisions under concurrent sensing-accuracy and communication-rate requirements. Buoy drift makes the topology time varying, while persistent queues couple current decisions to future sensing and service opportunities, posing a long-horizon combinatorial--continuous coordination challenge.

\subsection{Contributions}\label{subsec:contrib}
To address these challenges, this paper considers HAP-assisted sparse cooperative UAV-ISAC for long-horizon maritime data collection and develops a graph-based MARL framework tailored to its coupled discrete--continuous decision structure. The resulting design targets urgency-weighted clearance of buoy backlogs while accounting for sensing-accuracy and uplink-rate requirements. The main contributions are summarized as follows.
\begin{itemize}[leftmargin=1.5em]
	\item {We develop a HAP-coordinated UAV-ISAC model for maritime data collection that integrates a correlated sea-patch field, sea-patch-conditioned Singer buoy dynamics, echo sensing with viewing-angle-dependent radar cross section (RCS) and sea-state-dependent clutter, and fused posterior Cram\'er--Rao bounds (PCRBs) within a three-phase superframe. Unlike formulations based on aggregate rate or sensing performance, the resulting long-horizon objective measures communication utility through the service delivered to urgency-weighted buoy queues and captures the coupling among association, mobility, sensing, and queue evolution.}
	\item {We design a structured graph policy for sparse HAP-coordinated cooperation. A heterogeneous encoder represents the variable maritime topology, while a sequential $b$-matching decoder samples variable-cardinality UAV--buoy associations that satisfy candidate-edge and endpoint-capacity constraints by construction. A multi-agent proximal policy optimization (MAPPO)-style, coordinator-assisted CTDE procedure then combines the HAP-side discrete policy with parameter-shared UAV actors for decentralized motion refinement and sensing-power control, using a centralized critic supplied with the complete causal Markov state.}
	\item We evaluate the design against randomized, deterministic, optimization-based, and learning-based benchmarks. The proposed policy achieves the best overall sensing--service balance under the default setting, while encoder, decoder, and updater ablations identify relational aggregation, variable-cardinality association, and proximal policy optimization (PPO) updates as complementary contributors. Without retraining, it maintains its lead as the network size doubles, the mission horizon increases fivefold, and wave height and sustained current increase.
\end{itemize}

\begin{figure}[!t]
	\centering
	\includegraphics[width=0.75\columnwidth]{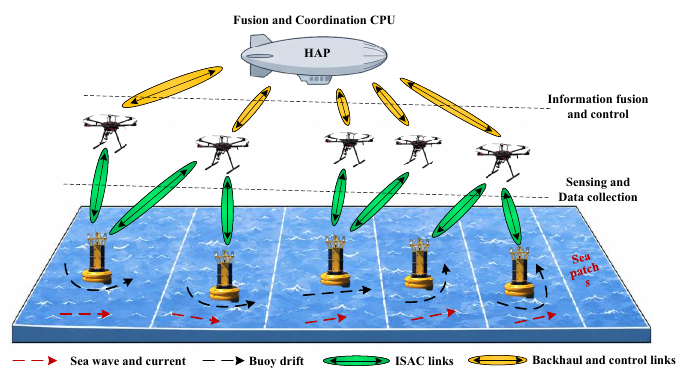}
	\vspace*{-2mm}
	\caption{System scenario of UAV-ISAC-assisted maritime data collection.}
	\label{fig:scenario}
\end{figure}\vspace*{-3mm}

\section{System Model}\label{sec:systemmodel}
\vspace*{-2mm}
We consider a HAP-assisted maritime network consisting of $K$ drifting buoys, $M$ rotary-wing fixed-height UAVs, and one HAP, as illustrated in Fig.~\ref{fig:scenario}. Each UAV is equipped with a uniform planar array (UPA) containing $R_2=R\times R$ antennas, while the HAP acts as a fusion-and-coordination central processing unit (CPU). In particular, the HAP receives local sensing and communication summaries from the UAVs, fuses buoy-state information, determines sparse cooperation patterns, and broadcasts high-level directives back to the UAVs\footnote{The considered compact HAP-UAV packets are assumed to be delivered reliably within the fusion phase. Control-link capacity, latency, and packet loss are outside the present model.}.

Let $T_{\mathrm m}$ be the mission duration, divided into $T$ superframes of duration $\Delta_{\mathrm T}=T_{\mathrm m}/T$. The system experiences three phases over superframe $t$, i.e.,
\begin{equation}\label{eq:superframe}
    \mathrm{S}_t \rightarrow \mathrm{U}_t \rightarrow \mathrm{F}_t,
\end{equation}
where $\mathrm{S}_t$, $\mathrm{U}_t$, and $\mathrm{F}_t$ denote the sensing, data collection, and fusion/control phases of superframe $t$, respectively. The UAV and buoy positions are assumed to be constant within one superframe\footnote{This quasi-static assumption applies to the short sensing and uplink sub-phases within a superframe, whereas buoy drift and UAV repositioning across superframes are modeled explicitly through the buoy dynamics in {(\ref{eq:buoydyn})} and the UAV mobility update in {(\ref{eq:positionupdate_revised})}.}.

At the beginning of $\mathrm{S}_t$, each UAV steers its sensing beam by using the shared one-step predicted buoy state from the previous superframe. The echo measurements collected during $\mathrm{S}_t$ are processed locally to form superframe-$t$ posterior estimates.
During $\mathrm{U}_t$, the UAVs receive uplink data from the assigned buoys. The uplink combiner used is constructed from the one-step predicted geometry. At the end of the uplink phase, UAV $m$ forms the packet as
\begin{align}\label{eq:zeta}
    \zeta_m[t]
    =
    \Big\{
        \hat{\pmb s}_{m,k}[t|t],\,
        \pmb P_{m,k}[t|t],\,
       \Gamma_{m,k}[t]
        %\,\iota_{m,k}^{\mathrm{int}}[t]
    \Big\}_{k\in\mathcal C_m[t]}
    \cup
    \Big\{\pmb c_m[t]\Big\},
\end{align}
where $\mathcal C_m[t]$ is the buoy set assigned to UAV $m$, $\hat{\pmb s}_{m,k}[t|t]$ and $\pmb P_{m,k}[t|t]$ are the local posterior state estimate and covariance of buoy $k$ after the {extended Kalman filter (EKF)} update \cite{ribeiro2004kalman}, $\Gamma_{m,k}[t]$ is the uplink post-combining signal-to-noise ratio (SNR) defined later in Section~\ref{subsec:fronthaul_uplink},
and $\pmb c_m[t]=[c_m^x[t],c_m^y[t],c_m^z]^{\mathrm T}$ is the UAV position with the fixed altitude.

During $\mathrm{F}_t$, the HAP collects $\{\zeta_m[t]\}_{m=1}^{M}$, fuses the uploaded posterior information, predicts the shared priors for the next superframe, and prepares for UAV $m$ the directive
\begin{equation}\label{eq:directive_msg}
\mathcal D_m[t+1]
=
\left\{
    \pmb q_m[t+1],\,
    \mathcal C_m[t+1],\,
    \pmb\nu_m^{\mathrm{dir}}[t+1]
\right\}.
\end{equation}
Here, $\pmb q_m[t+1]$ is a HAP-generated reference waypoint. The set $\mathcal C_m[t+1]$ is the next-superframe sparse UAV-buoy cluster, and $\pmb\nu_m^{\mathrm{dir}}[t+1]$ is a short directive-context vector used by the local UAV actor, which is defined explicitly in Section~\ref{subsec:actor}.

\subsection{Sea-Patch Correlated Maritime Field}\label{subsec:patch}
To represent spatially varying and time-varying sea conditions, the monitored square region of side length $L$ is partitioned into $S$ local sea patches indexed by $\mathcal S\triangleq\{1,\ldots,S\}$. Let $\pmb r_s\in\R^3$ denote the center of patch $s$. The environmental state of patch $s$ is defined as $\pmb z_s[t]=\left[H_s[t], \omega_{s}[t], v_{s,x}^{\text{cur}}[t], v_{s,y}^{\text{cur}}[t],\delta_s[t]\right]^{\mathrm T}$. Here, $H_s[t]$ is the significant wave height, $\omega_{s}[t]$ is the dominant wave angular frequency, $\pmb v_s^{\text{cur}}[t]\triangleq[v_{s,x}^{\text{cur}}[t],v_{s,y}^{\text{cur}}[t],0]^{\mathrm T}$ is the current velocity, and $\delta_s[t]$ is a dimensionless local clutter-perturbation state. The variables $H_s[t]$ and $\omega_{s}[t]$ represent local sea conditions and wave time scales in ocean-wave modeling \cite{Holthuijsen2007}, which directly affect buoy motion and clutter severity.

The inter-patch dependence is described by a directed graph $\mathcal G_{\text{sea}}[t]=(\mathcal S,\mathcal E_{\text{sea}},\pmb W[t])$,
where $\mathcal E_{\text{sea}}\subseteq\mathcal S\times\mathcal S$ is the directed edge set, and $\pmb W[t]=[w_{s,s^{\prime}}[t]]\in\R^{S\times S}$ is the time-varying weighted adjacency matrix. The ordered pair $(s,s^{\prime})\in\mathcal E_{\text{sea}}$ means that patch $s^{\prime}$ can transport environmental influence to patch $s$ over one superframe, and hence $w_{s,s^{\prime}}[t]$ is defined as the directed influence weight from source patch $s^{\prime}$ to destination patch $s$. For adjacent patches $s^{\prime}$ and $s$, we define the normalized current-aligned influence weight as \cite{hanert2004advection}
\begin{equation}\label{eq:patchweight}
	w_{s,s^{\prime}}[t]
	=
	\frac{
		e_{s,s^{\prime}}
		\exp\!\left(-\frac{\|\pmb r_s-\pmb r_{s^{\prime}}\|^2}{\sigma_d^2}\right)
		\psi_{s,s^{\prime}}[t]
	}{
		\sum_{\ell\in\mathcal N_{s^{\prime}}^{\text{out}}}
		e_{\ell, s^{\prime}}
		\exp\!\left(-\frac{\|\pmb r_{\ell}-\pmb r_{s^{\prime}}\|^2}{\sigma_d^2}\right)
		\psi_{\ell, s^{\prime}}[t]+\epsilon
	},
\end{equation}
where $e_{s,s^{\prime}}\in\{0,1\}$ equals one if $(s,s^{\prime})\in\mathcal E_{\text{sea}}$ and zero otherwise, $\mathcal N_{s^{\prime}}^{\text{out}}\triangleq\{\ell:(\ell,s^{\prime})\in\mathcal E_{\text{sea}}\}$ is the set of neighboring destination patches that can receive transport influence from source patch $s^{\prime}$, $\sigma_d^2>0$ is a spatial coupling-width parameter controlling the distance decay of inter-patch influence, and
\begin{equation}\label{eq:patchtransport}
	\psi_{s,s^{\prime}}[t]
	=\left[
	\frac{
		(\pmb r_s-\pmb r_{s^{\prime}})^{\mathrm T}\pmb v_{s^{\prime}}^{\text{cur}}[t]
	}{
		\|\pmb r_s-\pmb r_{s^{\prime}}\|\,\|\pmb v_{s^{\prime}}^{\text{cur}}[t]\|+\epsilon
	}
	\right]_+
\end{equation}
is the directional transport factor, where $[x]_+\triangleq\max\{x,0\}$.
The factor $\psi_{s,s^{\prime}}[t]$ quantifies how strongly the local current at source patch $s^{\prime}$ supports directional transport from patch $s^{\prime}$ to patch $s$. In particular, a larger $\psi_{s,s^{\prime}}[t]$ indicates stronger downstream transport from $s^{\prime}$ to $s$, whereas $\psi_{s,s^{\prime}}[t]=0$ means that the local current does not contribute to transport toward patch $s$. The small constant $\epsilon>0$ in \eqref{eq:patchweight} and \eqref{eq:patchtransport} is used only for numerical regularization when the denominator or the current norm becomes very small. Hence, nearby patches are more strongly coupled, while the coupling is reinforced along the local current direction.

The patch field then evolves according to\footnote{Equation~{(\ref{eq:patchdyn})} is a reduced-order stochastic closure that retains the short-horizon persistence, current-aligned transport, spatial locality, and unresolved process uncertainty needed by the UAV decision process.}
\begin{equation}\label{eq:patchdyn}
	\pmb z_s[t+1]
	=
	\pmb A_0\pmb z_s[t]
	+
	\sum_{s^{\prime}\in\mathcal N_s^{\text{in}}}
	w_{s,{s^{\prime}}}[t]\pmb A_1\pmb z_{s^{\prime}}[t]
	+
	\pmb n_s[t],
\end{equation}
where $\mathcal N_s^{\text{in}}\triangleq\{s^{\prime}:(s,s^{\prime})\in\mathcal E_{\text{sea}}\}$ is the set of source patches that can transport environmental influence to destination patch $s$, and $\pmb n_s[t]\sim\mathcal N(\pmb 0, \pmb Q_s)$ is the environmental process noise with covariance $\pmb Q_s=\operatorname{diag}\left(\sigma_{H, s}^2, \sigma_{\omega, s}^2, \sigma_{v_x, s}^2, \sigma_{v_y, s}^2, \sigma_{\delta, s}^2\right)$. We further define $\pmb A_0=\diag(\alpha_H,\alpha_{\omega},\alpha_v,\alpha_v,\alpha_{\delta})$ and $\pmb A_1=
\diag(\beta_H,\beta_{\omega},\beta_v,\beta_v,\beta_{\delta})$,
where $\pmb A_0$ captures the self-memory of each patch, whereas $\pmb A_1$ captures the neighbor-induced perturbation. The coefficients satisfy $0\le \alpha_q<1$, $0\le \beta_q<1$, and $\alpha_q+\beta_q\le 1$ for each environmental component $q\in\{H,\omega,v,\delta\}$.

Let $\rho_{k,s}[t]\in[0,1]$ be the interpolation coefficient from patch $s$ to buoy $k$, with $\sum_{s=1}^{S}\rho_{k,s}[t]=1$. The local sea condition experienced by buoy $k$ is then modeled as
\begin{align}\label{eq:buoypatchinterp}
\pmb z_k[t]
=\sum_{s=1}^{S}\rho_{k,s}[t]\pmb z_s[t]
=\big[H_k[t],\omega_k[t],v_{k,x}^{\mathrm{cur}}[t],
 v_{k,y}^{\mathrm{cur}}[t],\delta_k[t]\big]^{\mathrm T},
\end{align}
where $H_k[t]\triangleq\sum_{s=1}^{S}\rho_{k,s}[t]H_s[t]$ denotes the local significant wave height experienced by buoy $k$, and $\delta_k[t]\triangleq\sum_{s=1}^{S}\rho_{k,s}[t]\delta_s[t]$ is its interpolated local clutter perturbation.
Through \eqref{eq:buoypatchinterp}, the sea-patch field determines the current-driven drift and wave parameters in the buoy dynamics and indexes the local clutter conditions used by the sensing model. The patch layer therefore links the evolving maritime environment to the subsequent sensing and control decisions.\vspace*{-2mm}

\subsection{Patch-Aware Buoy Dynamics}\label{subsec:buoydyn}
The 3D position of buoy $k$ is denoted by $\pmb c_k[t]=[c_k^x[t],c_k^y[t],c_k^z]^{\mathrm T}$ and 
the horizontal motion state of buoy $k$ at superframe $t$ is given by\footnote{In this work, only the horizontal drift of each buoy is explicitly modeled, while its vertical position is not tracked.}
\begin{equation}\label{eq:buoystate}
	\pmb s_k[t]
	=\left[c_k^x[t],\bar v_k^x[t],a_k^x[t],c_k^y[t],\bar v_k^y[t],a_k^y[t]\right]^{\mathrm T},
\end{equation}
where $c_k^x[t]$ and $c_k^y[t]$ are the buoy coordinates, $\bar v_k^x[t]$ and $\bar v_k^y[t]$ are the wave-induced velocity components excluding the current drift, and $a_k^x[t]$ and $a_k^y[t]$ are the corresponding wave-induced acceleration components.

The short-term horizontal motion of a surface buoy is decomposed into a slowly varying current-induced drift and a wave-induced random fluctuation. The former is described by the interpolated local current velocity in \eqref{eq:buoypatchinterp}, whereas the latter is modeled by a Singer correlated-acceleration process \cite{Singer1970Tracking}. The resulting discrete-time state evolution is written as
\begin{equation}\label{eq:buoydyn}
	\pmb s_k[t+1]
	=
	\pmb G_k[t]\pmb s_k[t]
	+
	\Delta_{\mathrm T}\pmb v_k^{\mathrm{cur}}[t]
	+\boldsymbol{\xi}_k[t],
\end{equation}
where $\pmb v_k^{\mathrm{cur}}[t]=\left[v_{k,x}^{\text{cur}}[t],0,0,v_{k,y}^{\text{cur}}[t],0,0\right]^{\mathrm T}$ and $\pmb G_k[t]=[\bar{\pmb G}_k[t],\pmb{0};\pmb{0},\bar{\pmb G}_k[t]]$
is the 2D state-transition matrix, with $\bar{\pmb G}_k[t]$ denoting the one-axis Singer transition block. In particular, let $\varsigma_k[t]$ denote the reciprocal of the acceleration correlation time in the Singer model. Since the dominant short-term excitation of a surface buoy is wave-induced and its time scale is governed by the dominant local wave component, we can set $\varsigma_k[t]=\omega_k[t]$, where $\omega_k[t]=\sum_{s=1}^{S}\rho_{k,s}[t]\omega_s[t]$ is the local dominant wave angular frequency induced from the sea-patch interpolation in \eqref{eq:buoypatchinterp}. Then, following \cite{Singer1970Tracking}, the standard Singer discretization yields $\bar{\pmb G}_k[t]=[1, 0, 0;\Delta_{\mathrm T}, 1, 0;\phi_{1,k}[t], \phi_{2,k}[t], \lambda_k[t]]^{\mathrm T}$,
where $\lambda_k[t]= e^{-\varsigma_k[t]\Delta_{\mathrm T}}$ is the one-step memory factor of the discretized wave-induced acceleration process, and
\begin{equation}\label{eq:phi12k}
	\phi_{1,k}[t]
	=
	\frac{\varsigma_k[t]\Delta_{\mathrm T}-1+\lambda_k[t]}{\varsigma_k^2[t]},\
	\phi_{2,k}[t]
	=
	\frac{1-\lambda_k[t]}{\varsigma_k[t]}.
\end{equation}
The process noise $\boldsymbol{\xi}_k[t]\sim\mathcal N(\pmb 0,\pmb Q_k[t])$ captures the residual wave-driven uncertainty after discretization. To tie its intensity to the local wave condition, we parameterize the acceleration standard deviation as $\sigma_{a,k}[t]=c_a\,\omega_k^2[t]H_k[t]$,
where $c_a>0$ is a dimensionless calibration coefficient. The resulting discrete-time process-noise covariance is block diagonal and is given by $\pmb Q_k[t]
=
\operatorname{blkdiag}(\bar{\pmb Q}_k[t],\bar{\pmb Q}_k[t])$,
where $\bar{\pmb Q}_k[t]$ is assembled from $\sigma_{a,k}[t]$ and the Singer parameters $(\varsigma_k[t],\lambda_k[t])$ and its closed-form entries are derived in Appendix~A of the supplementary material.

\subsection{RCS-Aware Multi-UAV Sensing under Sea Clutter}\label{subsec:sensing}
For UAV $m$ and buoy $k$, we define the transmit/receive array response vectors as $\pmb \alpha_{m,k}^{\text{tx/rx}}[t]=\frac{1}{R}\,
\pmb \alpha_{m,k}^{x}[t]\otimes \pmb \alpha_{m,k}^{y}[t]$,
with
\begin{equation}
	\pmb \alpha_{m,k}^{x/y}[t]
	=
	\begin{bmatrix}
		1, e^{-j\mu_{m,k}^{x/y}[t]}, \cdots, e^{-j(R-1)\mu_{m,k}^{x/y}[t]}
	\end{bmatrix}^{\mathrm T},
\end{equation}
where
\begin{equation}\label{eq:array_phase}
	\mu_{m,k}^{x}[t]
	=
	\frac{2\pi d_{\text a}}{\lambda}
	\frac{c_m^x[t]-c_k^x[t]}{d_{m,k}[t]},\
	\mu_{m,k}^{y}[t]
	=
	\frac{2\pi d_{\text a}}{\lambda}
	\frac{c_m^y[t]-c_k^y[t]}{d_{m,k}[t]}.
\end{equation}
Here, $d_{m,k}[t]=\|\pmb c_m[t]-\pmb c_k[t]\|$, while $d_{\text a}$ and $\lambda$ denote the antenna spacing and carrier wavelength, respectively.
%with as $\pmb c_m[t]$ defined as $\pmb c_m[t]=[\bar{\pmb c}_m^{\mathrm T}[t], c^z_m]^{\mathrm T}$,

During $\mathrm{S}_t$, UAV $m$ allocates power $p_{m,k}^{\mathrm S}[t]$ to sensing buoy $k$ in its assigned buoy set $k\in\mathcal C_m[t]$, subject to the fixed sensing-power budget
\begin{equation}\label{eq:powerbudget}
	\sum_{k\in\mathcal C_m[t]}p_{m,k}^{\mathrm S}[t]
	\le P_m^{\max},\ \forall m,t.
\end{equation}
Then, the sensing signal transmitted by UAV $m$ during $\mathrm{S}_t$ is
\begin{equation}\label{eq:sensing_tx}
	\pmb x_m^{\mathrm S}[t]
	=
	\sum_{k=1}^K
	\gamma_{m,k}[t]\sqrt{p_{m,k}^{\mathrm S}[t]}\,
	\pmb f_{m,k}[t]\,
	x_{m,k}^{\text S}[t],
\end{equation}
where the binary cluster indicator $\gamma_{m,k}[t]\in\{0,1\}$ denotes the association between the UAV $m$ and buoy $k$, $\pmb f_{m,k}[t]\in\mathbb C^{R_2\times 1}$ is the transmit beamformer and $x_{m,k}^{\mathrm S}[t]$ is the unit-power baseband sensing signal with $\EE[|x_{m,k}^{\text S}[t]|^2]=1$. 
Since the UAV-buoy links are dominated by LoS propagation and the sensing beam is steered according to the one-step prediction, we set $\pmb f_{m,k}[t]=\hat{\pmb a}_{m,k}^{\text{tx}}[t|t-1]$. As the UAV is equipped with a large MIMO array, its narrow pencil beam can effectively suppress inter-target interference at the receiver \cite{dong2022sensing}. Moreover, maritime buoys are generally sparsely deployed, providing clearer angular separation than typical terrestrial targets.
The local sensing echo associated with buoy $k$ is modeled as
\begin{align}\label{eq:localsense}
	&\pmb r_{m,k}^{\text S}[t]
	=
	\sqrt{p_{m,k}^{\text S}[t]}
	\Big(
	\beta_{m,k}[t]\pmb \alpha_{m,k}^{\text{rx}}[t](\pmb \alpha_{m,k}^{\text{tx}}[t])^{\text H}+\varrho_{m,k}^{\mathrm{clu}}[t]\times
	\nonumber\\
	&\beta_{m,k}^{\text{clu}}[t] \hat{\pmb a}_{m,k}^{\text{rx}}[t|t-1](\hat{\pmb a}_{m,k}^{\text{tx}}[t|t-1])^{\text H}
	\Big)\pmb f_{m,k}[t]
	+
	\pmb n_{m,k}[t],
\end{align}
where $\pmb \alpha_{m,k}^{\mathrm{tx}}[t]$ and $\pmb \alpha_{m,k}^{\mathrm{rx}}[t]$ follow the realized LoS geometry, whereas $\hat{\pmb a}_{m,k}^{\mathrm{tx}}[t|t-1]$ and $\hat{\pmb a}_{m,k}^{\mathrm{rx}}[t|t-1]$ are evaluated at the one-step predicted buoy position $\hat{\pmb c}_k[t|t-1]$. The beamformer and clutter look direction are therefore determined causally before $\mathrm S_t$, while the target echo depends on the realized buoy position. Moreover, $\beta_{m,k}[t]$ is the target reflection coefficient, $\beta_{m,k}^{\mathrm{clu}}[t]$ is the residual sea-clutter reflection coefficient, $\varrho_{m,k}^{\mathrm{clu}}[t]\in(0,1)$ models residual clutter leakage after front-end suppression, and $\pmb n_{m,k}[t]\sim\mathcal{CN}(\pmb 0,N_0\pmb I_{R_2})$ is the receiver noise, with $N_0$ denoting the noise power integrated over bandwidth $B$.

The target reflection coefficient $\beta_{m,k}[t]$ follows an RCS-aware buoy model. Buoy $k$ is treated as a quasi-upright symmetric target with a single effective scattering center. Following the general single-scattering-point factorization in \cite{etsi_isc002_2025,le2025isacrcs}, the RCS contains a buoy-type reference term, a deterministic viewing-angle factor, and a fast-varying factor. {In the simulations, the fast-varying factor is fixed to one, so $\beta_{m,k}[t]$ is determined by the buoy-specific reference RCS, viewing geometry, and propagation distance.} The residual sea-clutter coefficient $\beta_{m,k}^{\mathrm{clu}}[t]$ is produced by the narrow-beam illuminated sea-surface patch around the predicted look direction, and its backscattering level is indexed by the local sea state of buoy $k$ \cite{GregersHansen2012SeaClutter}. Closed-form expressions for both coefficients are provided in Appendix~B of the supplementary material.\vspace*{-2mm}

\subsection{HAP-Fused PCRB Metric}\label{subsec:pcrb}
The common predicted prior and the local echoes collected in superframe $t$ provide the information used by the cooperative sensing metric. After matched filtering the raw sensing echo $\pmb r_{m,k}^{\mathrm S}[t]$ in \eqref{eq:localsense}, the measurement model is
\begin{equation}\label{eq:compact_meas}
	\pmb m_{m,k}[t]
	=
	\bar{\pmb r}_{m,k}\!\big(\pmb s_k[t]\big)
	+
	\tilde{\pmb n}_{1,m,k}[t],
\end{equation}
where the informative echo mean is
$\bar{\pmb r}_{m,k}\!\big(\pmb s_k[t]\big) =
	\beta_{m,k}[t]\varpi_{m,k}^{\mathrm{tx}}[t]
	\pmb\alpha_{m,k}^{\mathrm{rx}}[t]$
with $\varpi_{m,k}^{\mathrm{tx}}[t]=\big(\pmb\alpha_{m,k}^{\mathrm{tx}}[t]\big)^{\mathrm H}\hat{\pmb a}_{m,k}^{\mathrm{tx}}[t|t-1]$.
The effective echo-noise block satisfies $\tilde{\pmb n}_{1,m,k}[t]
\sim
\mathcal{CN}\!\big(
\pmb 0,\,
\sigma_{1,m,k}^2[t]\pmb I_{R_2}
\big)$,
where $\sigma_{1,m,k}^2[t]={c_1^2N_0}/{p_{m,k}^{\mathrm S}[t]M_f^2}+c_1^2|\varrho_{m,k}^{\mathrm{clu}}[t]\beta_{m,k}^{\mathrm{clu}}[t]|^2$, $M_f$ denotes the deterministic matched-filter amplitude gain and $c_1$ is a constant related to the signal design and system configuration\footnote{After front-end clutter suppression and matched filtering, the
unresolved residual clutter is modeled as a zero-mean spatially white disturbance that is independent of the thermal noise. Accordingly, the two components are absorbed into a single effective echo-noise variance.}.
Linearizing \eqref{eq:compact_meas} at the common prior mean $\hat{\pmb s}_k[t|t-1]$ yields
\begin{align}\label{eq:local_linearization}
	\pmb m_{m,k}[t]
	&\approx
	\bar{\pmb r}_{m,k}\!\big(\hat{\pmb s}_k[t|t-1]\big)
	+
	\pmb H_{m,k}[t]
	\big(
	\pmb s_k[t]-\hat{\pmb s}_k[t|t-1]
	\big)\nonumber\\
	&\quad+
	\tilde{\pmb n}_{1,m,k}[t],
\end{align}
with $\pmb H_{m,k}[t]=\left.
\frac{\partial \bar{\pmb r}_{m,k}(\pmb s)}
{\partial \pmb s^{\mathrm T}}
\right|_{\pmb s=\hat{\pmb s}_k[t|t-1]}$.
Under this linearized complex Gaussian model, the local Fisher information matrix (FIM) contributed by UAV $m$ is derived as
\begin{equation}\label{eq:local_fim}
	\pmb J_{m,k}^{\text{loc}}[t]
	=
	\frac{2}{\sigma_{1,m,k}^2[t]}
	\Re\!\Big\{
	\pmb H_{m,k}^{\text H}[t]
	\pmb H_{m,k}[t]
	\Big\}.
\end{equation}

Combining the common prior with the local information from all active sensing UAVs, the HAP forms the fused FIM as
\begin{equation}\label{eq:fusedfim}
	\pmb J_k[t]
	=
	\pmb P_k^{-1}[t|t-1]
	+
	\sum_{m=1}^{M}
	\gamma_{m,k}[t]\pmb J_{m,k}^{\text{loc}}[t],
\end{equation}
where $\pmb P_k[t|t-1]$ is the common one-step prior covariance shared by all sensing UAVs before the superframe-$t$ echoes are processed. It is generated by the echo-only EKF prediction from the previous fused posterior pair $(\hat{\pmb s}_k[t-1|t-1],\pmb P_k[t-1|t-1])$.
The complete local-update, HAP-fusion and next-step-prediction recursion used to obtain $\pmb P_k[t|t-1]$, $\pmb P_k[t|t]$, and $\hat{\pmb s}_k[t|t]$ is given in Appendix~C of the supplementary material.
The corresponding linearized fused horizontal-position PCRB is used as the sensing metric of buoy $k$ and is given by
\begin{equation}\label{eq:pcrb}
	\Theta_k[t]
	=
	\tr\!\left(
	\pmb E_{\text p}^{\mathrm T}
	\pmb J_k^{-1}[t]
	\pmb E_{\text p}
	\right),
\end{equation}
where $\pmb E_{\text p}=[1,0,0,0,0,0;0,0,0,1,0,0]^{\text T}$.
Note that the multi-UAV sensing metric is cluster coupled. When several UAVs observe the same buoy, their information contributions combine through \eqref{eq:fusedfim}--\eqref{eq:pcrb}, so the fused sensing quality depends jointly on the collective geometry, local RCS and clutter conditions, and the participating UAV set.

\subsection{HAP-Coordinated Sparse Uplink Communication}\label{subsec:fronthaul_uplink}
At $\mathrm{U}_t$, the uplink access channel between buoy $k$ and UAV $m$ is modeled as $\pmb h_{m,k}^{\mathrm U}[t]
=\sqrt{\beta_{m,k}^{\mathrm U}[t]}\,\pmb \alpha_{m,k}^{\text {rx}}[t]$,
where $\beta_{m,k}^{\mathrm U}[t]=\beta_0 {d}_{m,k}^{-\alpha_{\text U}}[t]$
is the large-scale uplink channel gain, with $\beta_0$ being the reference gain and $\alpha_{\text U}$ the access-link path-loss exponent.

Let $x_{m,k}^{\mathrm U}[t]$ denote the uplink data transmitted from buoy $k$ to UAV $m$.
The local received signal at UAV $m$ is then derived as\footnote{At buoy $k$, it is briefly assumed that the data streams for different UAVs are mapped upon mutually orthogonal subbands. Also, similar to the sensing echo receiving, the inter-buoy interference is neglected owing to narrow beams and sparse distribution of buoys.}
\begin{equation}\label{eq:local_uplink}
	y_{m,k}^{\mathrm U}[t]
	=
	\pmb w_{m,k}^{\text H}[t]
	\Big(
	\sqrt{P_k^{\mathrm U}}\pmb h_{m,k}^{\mathrm U}[t]x_{m,k}^{\text U}[t]
	+
	\pmb n_m^{\mathrm U}[t]
	\Big),
\end{equation}
where $P_k^{\mathrm U}$ is transmit power, $\pmb n_m^{\mathrm U}[t]\sim\mathcal{CN}(\pmb 0,N_0\pmb I_{R_2})$ is the uplink noise, and the receive combiner is set as $\pmb w_{m,k}[t]=\hat{\pmb a}_{m,k}^{\text{rx}}[t|t-1]$. The corresponding local post-combining SNR with respect to $k$ at UAV $m$ is
\begin{equation}\label{eq:local_snr}
	\Gamma_{m,k}[t]
	=
	{
		P_k^{\mathrm U}
		\big|
		\pmb w_{m,k}^{\text H}[t]\pmb h_{m,k}^{\mathrm U}[t]
		\big|^2
	}/{
	N_0\|\pmb w_{m,k}[t]\|^2
	}.
\end{equation}
Hence, the aggregate uplink rate associated with buoy $k$ is modeled as
\begin{equation}\label{eq:rate}
	R_k[t]=
	\sum_{m=1}^M
	\gamma_{m,k}[t]\log_2\!\left(1+\Gamma_{m,k}[t]\right).
\end{equation}

\subsection{Queue-Weighted Buffered Collection State}\label{subsec:queue}
The optimization target is not instantaneous uplink throughput alone, but the timely clearance of buffered maritime observations across multiple superframes. We therefore model a data queue at each buoy so that service in one superframe determines both current collection and the backlog carried into later decisions.
Specifically, at superframe $t$, buoy $k$ carries a traffic state consisting of a backlog $b_k[t]\ge 0$ of buffered observations, an exogenous arrival $a_k[t]\ge 0$, and an urgency weight $u_k\ge 0$.
The per-edge uplink rate is converted into a rate-coupled service capacity as $s_{m,k}[t]=\alpha_R\,R^{\mathrm U}_{m,k}[t]$,
where $R^{\mathrm U}_{m,k}[t]=\log_2(1+\Gamma_{m,k}[t])$ is the realized per-edge uplink rate, and $\alpha_R>0$ maps link quality to served data units. The available and the collected data at buoy $k$ are given by $A_k[t]=b_k[t]+a_k[t]$, and
\begin{align}\label{eq:queue_service}
C_k[t]=\min\!\Big(A_k[t],\ \sum_{m=1}^{M}\gamma_{m,k}[t]\,s_{m,k}[t]\Big),
\end{align}
respectively, so that collection is jointly capped by the available backlog and by the cluster service capacity. The backlog then recurses as
\begin{equation}\label{eq:queue_recursion}
b_k[t+1]=\big[\,b_k[t]+a_k[t]-C_k[t]\,\big]_+,
\end{equation}
with $[x]_+\triangleq\max\{x,0\}$. Equations~\eqref{eq:queue_service}--\eqref{eq:queue_recursion} connect communication quality to long-horizon control. A higher realized uplink rate increases service capacity and can therefore increase current collection while reducing subsequent backlog. %Hence, a high-rate UAV-buoy link is valuable only when it is selected and the served buoy carries useful backlog, so the resulting design is rate-aware rather than pure-rate maximizing.
\vspace*{-2mm}

\section{Problem Formulation}\label{sec:problem}
The system task is to coordinate UAV--buoy association, UAV motion, and sensing-power allocation over the mission horizon so as to clear buffered buoy observations in a timely and urgency-aware manner. To reflect this task, the reward combines the fraction of newly available data collected, urgency-weighted and backlog-weighted service, and the positive one-step reduction of a quadratic backlog potential. Since the raw magnitudes of these terms vary with the traffic realization and queue scale, it is essential to normalize them into bounded, dimensionless forms.

To this end, before defining the reward components,
we introduce the superframe-wise normalizers as
\noindent\scalebox{0.9}{$S_A[t]\triangleq\max\{\sum_{k=1}^{K}A_k[t],\,1\}, S_U[t]\triangleq\max\{2\sum_{k=1}^{K}u_kA_k[t],\,1\}$}, \scalebox{0.9}{$B_{\max}[t]\triangleq \max\{\max_k b_k[t],\,1\}$}, and $\Phi(\pmb b[t])\triangleq \sum_{k=1}^{K}b_k^2[t]$,
where $\pmb b[t]=[b_1[t],\ldots,b_K[t]]^{\mathrm T}$ denotes the backlog vector, $S_A[t]$ is the available-data scale, $S_U[t]$ is the urgency-weighted service scale, and $B_{\max}[t]$ measures the instantaneous queue-pressure scale.
The three reward components are then defined as
\begin{align}
&r_{\mathrm{data}}[t]
={\sum_{k=1}^{K} C_k[t]}/{S_A[t]},\nonumber \\
&r_{\mathrm{queue}}[t]
={\sum_{k=1}^{K} u_k C_k[t]\left(1+{b_k[t]}/{B_{\max}[t]}\right)}/{S_U[t]}, \nonumber \\
&r_{\mathrm{pot}}[t]
=\left[(\Phi(\pmb b[t])-\Phi(\pmb b[t+1]))/{\max\{\Phi(\pmb b[t]),1\}}\right]_+.
\label{eq:r_potential}
\end{align}
Thus, \(r_{\mathrm{pot}}[t]\) is positive only when the quadratic backlog potential actually decreases from one superframe to the next.

Finally, the per-superframe queue-weighted utility is given by $r_Q[t]
=
w_{\mathrm{data}}r_{\mathrm{data}}[t]
+w_{\mathrm{queue}}r_{\mathrm{queue}}[t]
+w_{\mathrm{pot}}r_{\mathrm{pot}}[t]$,
Here, $r_{\mathrm{data}}[t]$ measures the fraction of available data collected in the current superframe, $r_{\mathrm{queue}}[t]$ weights current service by urgency and backlog pressure, and $r_{\mathrm{pot}}[t]$ rewards a positive one-step decrease of the quadratic backlog potential. Thus, $r_Q[t]$ combines immediate collection, priority-aware service, and backlog reduction rather than representing network throughput alone.

Using this utility, the long-horizon control problem maximizes the timely and urgency-weighted clearance of buffered buoy observations subject to sensing, communication, mobility, and sparse-cooperation requirements. It is formulated as
\begin{subequations}\label{prob:P1}
	\begin{align}
		(\mathrm{P1}):\quad
		&\max_{\{\pmb c_m[t],\gamma_{m,k}[t],p_{m,k}^{\mathrm S}[t]\}}\mathbb{E}\left[\sum_{t=1}^{T} r_Q[t]\right]
		\label{prob:P1obj}\\
		\mathrm{s.t.}
		& \sum_{k\in\mathcal C_m[t]}p_{m,k}^{\mathrm S}[t]
		\le P_m^{\max},\
		\forall m,t,
		\label{prob:P1pow}\\
		& \sum_{m=1}^{M}\gamma_{m,k}[t]
		\le L_{\max},\
		\forall k,t,
		\label{prob:P1cluster}\\
		& \sum_{k=1}^{K}\gamma_{m,k}[t]
		\le D_{\max},\
		\forall m,t,
		\label{prob:P1load}\\
		& \gamma_{m,k}[t]\in\{0,1\},\
		\forall m,k,t,
		\label{prob:P1binary}\\
		& \|\pmb c_m[t]-\pmb c_m[t-1]\|
		\le V_{\max}\Delta_{\mathrm T},\
		\forall m,t,
		\label{prob:P1move}\\
		& 0\le v_m[t]\le V_{\max},\
		\forall m,t,
		\label{prob:P1speedbound}\\
		& \Theta_k[t]\le \Theta_{\max},\
		\forall k,t,
		\label{prob:P1sense}\\
		& R_k[t]\ge R_{\min}{\,\bar\gamma_k[t]},\
		\forall k,t,
		\label{prob:P1rate}
	\end{align}
\end{subequations}
where {$v_m[t]\triangleq\|\pmb c_m[t]-\pmb c_m[t-1]\|/\Delta_{\mathrm T}$ is the average UAV speed over superframe $t$}, $L_{\max}$ and $D_{\max}$ are the maximum buoy-cluster size and UAV serving load, respectively, $V_{\max}$ is the maximum UAV speed, $\Theta_{\max}$ is the allowed horizontal-position PCRB threshold, and $R_{\min}$ is the required uplink-rate threshold{, while $\bar\gamma_k[t]\in\{0,1\}$ indicates whether buoy $k$ is served by at least one UAV in superframe $t$}. Constraint \eqref{prob:P1pow} imposes the sensing-power budget, \eqref{prob:P1cluster}--\eqref{prob:P1binary} define sparse feasible cooperation, \eqref{prob:P1move}--\eqref{prob:P1speedbound} impose mobility consistency, and \eqref{prob:P1sense}--\eqref{prob:P1rate} impose sensing and communication quality requirements. Note that the minimum-rate requirement in \eqref{prob:P1rate} is imposed
only on actively served buoys, that is, when a buoy is not
selected in a superframe, the missing service is accounted for by the backlog recursion rather than by an infeasible per-slot rate guarantee.

\begin{remark}
	At the decision epoch, the HAP and the UAVs act only on one-step predictions of the buoy states and link qualities. After the association and the local motion actions are executed, the service $s_{m,k}[t]$, the collection $C_k[t]$, the backlog $b_k[t+1]$, and the utility $r_Q[t]$ are evaluated with the realized per-edge rate $R^{\mathrm U}_{m,k}[t]$. In other words, predictions guide decisions, whereas realizations drive the physical service and the queue evolution.
\end{remark}

We should also mention that it is impractical to exactly guarantee the rate and PCRB constraints in learning-based optimization methods \cite{Jiang2026HADMARL}. Hence, they are transformed into the normalized soft costs during training, which are given by
\begin{align}
 g_{\Theta,k}[t]&=\left[{\Theta_k[t]}/{{\Theta_{\max}}}-1\right]_+,\label{eq:g_theta_rw_revised}\\
 g_{R,k}[t]&={[R_{\min}{\bar\gamma_k[t]}-R_k[t]]_+}/{\max\{R_{\min},\varepsilon_R\}},\label{eq:g_rate_rw_revised}
\end{align}
where $\varepsilon_R>0$ is a small numerical floor. {The resulting training reward is}
\begin{equation}\label{eq:reward_revised}
\begin{aligned}
r[t]
=&r_Q[t]
-\lambda_{\Theta}^{\mathrm{rw}}\sum_k g_{\Theta,k}[t]
-\lambda_R^{\mathrm{rw}}\sum_k g_{R,k}[t],
\end{aligned}
\end{equation}
with nonnegative weights $(\lambda_{\Theta}^{\mathrm{rw}},\lambda_R^{\mathrm{rw}})$.

\section{Structured Feasible-Preserving Graph-MARL}\label{sec:algorithm}
\begin{figure*}[!t]
	\centering
	\includegraphics[width=0.95\textwidth]{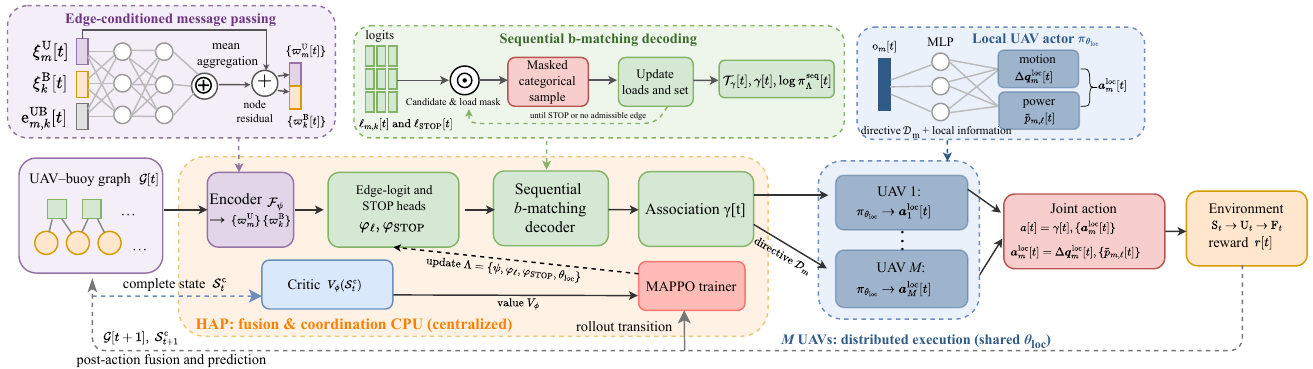}
	\vspace*{-2mm}
	\caption{{Overall architecture and CTDE training flow of the proposed SFPG-MARL framework.}}
	\label{fig:model}
\end{figure*}

The main algorithmic difficulty in (P1) is coordinating sparse binary association with continuous UAV control. The binary association matrix $\gamma[t]\triangleq[\gamma_{m,k}[t]]\in\{0,1\}^{M\times K}$ must respect the candidate mask, the per-UAV load limit, as well as the per-buoy cluster limit, and every selected edge simultaneously changes the sensing quality, the rate-coupled service, the backlog evolution, and the future graph state.
We therefore construct a structured stochastic policy that samples a feasible ordered association trace, maps it to a binary association, and evaluates its trace log-probability for PPO updates. The remainder of this section first presents the coordinator-assisted decentralized partially observable Markov decision process (Dec-POMDP) and its causal coordination graph, then develops the feasibility-preserving $b$-matching policy and the corresponding local-control realization, and finally describes centralized-critic PPO training. Fig.~\ref{fig:model} summarizes the resulting structured feasible-preserving graph-MARL (SFPG-MARL) architecture and its CTDE information flow.

\subsection{Coordinator-Assisted Dec-POMDP and Heterogeneous Graph}\label{subsec:decpomdp_control}
In the coordinator-assisted Dec-POMDP, the HAP makes the globally coupled association decision, whereas the UAVs share the mission reward and execute local continuous control from their directives and partial observations. Before superframe $t$ begins, the HAP selects $\gamma[t]$ from $\mathcal G[t]$, which induces $\mathcal C_m[t]\triangleq\{k\mid\gamma_{m,k}[t]=1\}$ for UAV $m$. Once $\mathcal C_m[t]$ is fixed, the parameter-shared UAV actor samples $\pmb a_m^{\mathrm{loc}}[t]$ from its directive and local observation. The resulting joint policy action is
\begin{equation}\label{eq:full_action_split}
 a[t]=\big(\gamma[t],\pmb a_1^{\mathrm{loc}}[t],\ldots,\pmb a_M^{\mathrm{loc}}[t]\big).
\end{equation}

Since the association is chosen before superframe $t$, $\mathcal G[t]$ contains only information available at that decision epoch. It uses the previous executed UAV and directive states, the current queue, arrival, and sea-patch descriptors, and the one-step buoy predictions. Accordingly, the HAP constructs the typed graph
as $\mathcal G[t]=\big(\mathcal V^{\mathrm U}\cup\mathcal V^{\mathrm B},\mathcal E[t]\big)$. The UAV node feature is given by
\begin{align}\label{eq:uavfeat_revised}
\pmb \xi_m^{\mathrm U}[t]
=\big[
\pmb c_m^{\mathrm T}[t-1]/L,\,
\pmb q_m^{\mathrm T}[t-1]/L,n_m^{\mathrm{cand}}[t]/K,\,
D_{\max}/K
\big]^{\mathrm T},
\end{align}
and the buoy node feature is designed as
\begin{align}\label{eq:buoyfeat_revised}
&\pmb \xi_k^{\mathrm B}[t]
=
\big[
\hat c_{k}^x[t|t-1]/L,\,\hat c_{k}^y[t|t-1]/L,\,
b_k[t]/B_{\max}[t], \nonumber \\
&a_k[t]/A_{\max}[t], u_k,\,
\widehat\Theta_k^{(0)}[t],\,\mathrm{rank}_k[t]/(K-1)
\big]^{\mathrm T},
\end{align}
where $A_{\max}[t]=\max\{\max_j a_j[t],1\}$, $\pmb q_m[t-1]$ is the directive waypoint executed in the previous superframe, and the division by the area side $L$ renders all position-type entries dimensionless. At mission initialization, the previous-waypoint convention is $\pmb q_m[0]=\pmb c_m[0]$. Furthermore, $n_m^{\mathrm{cand}}[t]\triangleq\sum_{k=1}^{K}\chi_{m,k}[t]$ is the UAV candidate load, with $\chi_{m,k}[t]\in\{0,1\}$ being the candidate-edge indicator.
$\hat c_{k}^x[t|t-1]$ and $\hat c_{k}^y[t|t-1]$ are the predicted buoy horizontal coordinates. In addition, $\widehat\Theta_k^{(0)}[t]$ is the prior horizontal PCRB, and $\mathrm{rank}_k[t]$ is the descending backlog rank of buoy $k$. The buoy queue summaries $(b_k,a_k,u_k)$ are deliberately exposed on both the buoy node and the candidate edge so that backlog and urgency inform both the node embeddings and the per-edge logits.

Based on the same predicted geometry, we define the causal pre-action screening signal-to-clutter-plus-noise ratio (SCNR) as
\begin{equation}\label{eq:candidate_scnr_revised}
\widehat{\mathrm{SCNR}}_{m,k}[t]
=\frac{P_m^{\max}|\widehat\beta_{m,k}[t]|^2}
{P_m^{\max}|\widehat\varrho_{m,k}^{\mathrm{clu}}[t]
\widehat\beta_{m,k}^{\mathrm{clu}}[t]|^2+N_0},
\end{equation}
where the full-power reference $P_m^{\max}$ makes the screening quantity independent of the sensing-power allocation subsequently selected for the retained association.

The UAV--buoy edge feature collects geometric, sensing, communication, queue, and load quantities and is defined as
\begin{align}\label{eq:ubedge_revised}
&\pmb e_{m,k}^{\mathrm{UB}}[t]
=\big[\chi_{m,k}[t],\,\hat d_{m,k}[t],
{\widehat{\mathrm{SCNR}}_{m,k}[t]},
\widehat R_{m,k}^{\mathrm U}[t], \nonumber\\
&\widehat{\Delta\Theta}_{m,k}[t],\bar b_k[t], u_k, \bar a_k[t],
\gamma_{m,k}[t-1], \nonumber\\
&n_m^{\mathrm{cand}}[t]/K,\,D_{\max}/K,\,
l_k^{\mathrm{cand}}[t]/M\big]^{\mathrm T},
\end{align}
where {$\hat d_{m,k}[t]=\|\pmb c_m[t-1]-\hat{\pmb c}_k[t|t-1]\|$}, $\bar b_k[t]\triangleq{b_k[t]}/{B_{\max}[t]}$, $\bar a_k[t]\triangleq{a_k[t]}/{\max\{\max_j a_j[t],1\}}$, and $l_k^{\mathrm{cand}}[t]\triangleq{\sum_{i=1}^{M}\chi_{i,k}[t]}$.
The queue-related entries $(\bar b_k[t],\bar a_k[t],u_k)$ make the sampled association backlog- and urgency-aware, while the load summaries expose local sparse-feasibility pressure.

Evaluating the echo mean and Jacobian at $\hat{\pmb s}_k[t|t-1]$ gives
\begin{align}
\widehat{\pmb J}_{m,k}^{\mathrm{nom}}[t]
={2}/{\hat\sigma_{1,m,k}^2[t]}
\Re\!\left\{\widehat{\pmb H}_{m,k}^{\mathrm H}[t]\widehat{\pmb H}_{m,k}[t]\right\},\label{eq:nominal_fim_revised}
\end{align}
where $\widehat{\pmb H}_{m,k}[t]$ uses the predicted buoy state and the pre-action UAV position $\pmb c_m[t-1]${, and $\hat\sigma_{1,m,k}^2[t]$ is the corresponding effective echo-noise variance evaluated at the predicted geometry}. The predicted marginal PCRB improvement is $\widehat{\Delta\Theta}_{m,k}[t]
=\big[\widehat\Theta_k^{(0)}[t]-\widehat\Theta_{m,k}^{(1)}[t]\big]_+$, where $\widehat\Theta_k^{(0)}[t]
=\tr\!\big(\pmb E_{\mathrm p}^{\mathrm T}\pmb P_k[t|t-1]\pmb E_{\mathrm p}\big)$ and $\widehat\Theta_{m,k}^{(1)}[t]
=\tr\!\big(
\pmb E_{\mathrm p}^{\mathrm T}
\big(\pmb P_k^{-1}[t|t-1]+\widehat{\pmb J}_{m,k}^{\mathrm{nom}}[t]\big)^{-1}
\pmb E_{\mathrm p}
\big)$.
Similarly, the per-edge rate proxy is given by $\widehat R^{\mathrm U}_{m,k}[t]
=\log_2\!\big(1+\widehat\Gamma_{m,k}[t]\big)$,
where $\widehat\Gamma_{m,k}[t]$ is the predicted counterpart of \eqref{eq:local_snr}, formed with the predicted channel $\widehat{\pmb h}^{\mathrm U}_{m,k}[t|t-1]$. Both $\widehat R^{\mathrm U}_{m,k}[t]$ and $\widehat{\Delta\Theta}_{m,k}[t]$ are computed before the current sensing and uplink phases. They are used as graph features and waypoint weights, whereas the realized $\Theta_k[t]$, $R_k[t]$, service, and reward are evaluated only after the selected association and local actions are executed.

The graph encoder is designed as $\big(\{\pmb \varpi_m^{\mathrm U}[t]\},\{\pmb \varpi_k^{\mathrm B}[t]\}\big)
=
\mathcal F_\psi(\mathcal G[t]),$
where $\mathcal F_\psi(\cdot)$ is a type-aware message-passing network that exchanges messages between the UAV and buoy nodes along the candidate edges. For every retained UAV-buoy pair, the edge-logit head outputs $\ell_{m,k}[t]
=f_\ell\!\left(
\pmb \varpi_m^{\mathrm U}[t],\pmb \varpi_k^{\mathrm B}[t],
\pmb e_{m,k}^{\mathrm{UB}}[t];\varphi_\ell
\right)$, which parameterizes an edge action. Since the appropriate number of associations varies with the current network state, the decoder must also determine when to terminate the sequential edge-selection process. Accordingly, a separate learned STOP logit is generated from the pooled graph representation as
\begin{equation}\label{eq:stop_logit_revised}
\ell_{\mathrm{STOP}}[t]
=f_{\mathrm{STOP}}\!\left(
\left[\bar{\pmb\varpi}^{\mathrm U}[t];
\bar{\pmb\varpi}^{\mathrm B}[t]\right];
\varphi_{\mathrm{STOP}}
\right).
\end{equation}
Here, $\bar{\pmb\varpi}^{\mathrm U}[t]=M^{-1}\sum_m\pmb\varpi_m^{\mathrm U}[t]$ and $\bar{\pmb\varpi}^{\mathrm B}[t]=K^{-1}\sum_k\pmb\varpi_k^{\mathrm B}[t]$.
Non-candidate pairs receive a mask value of $-\infty$ before decoding and never enter the sampled action.\vspace*{-2mm}

\subsection{Feasibility-Preserving Stochastic $b$-Matching Policy}\label{subsec:projection}
A $b$-matching is a matching on a graph in which each vertex has an integer capacity $b$ rather than the unit capacity used in ordinary matching. In the UAV-buoy bipartite graph, the UAV-side capacities are $b_m=D_{\max}$ and the buoy-side capacities are $b_k=L_{\max}$.
The decoder operates on the candidate graph defined by the {distance-and-SCNR mask}
$\chi_{m,k}[t]
=\mathbf 1\!\left\{
\hat d_{m,k}[t]\le d_{\mathrm{cand}},\;
{\widehat{\mathrm{SCNR}}_{m,k}[t]\ge\Gamma_{\mathrm{cand}}}
\right\}$,
with fixed thresholds $d_{\mathrm{cand}}$ and {$\Gamma_{\mathrm{cand}}$}. The retained edge set is $\mathcal E_{\mathrm{cand}}[t]=\{(m,k):\chi_{m,k}[t]=1\}.$
The feasible association family is the set of all binary capacity-constrained $b$-matchings on the candidate graph, given by
\begin{align}\label{eq:feasible_family_revised}
&\mathcal F_\gamma[t]=\big\{\gamma\in\{0,1\}^{M\times K}:\;
\gamma_{m,k}[t]\le \chi_{m,k}[t],\;\forall m,k,\nonumber \\
&\sum_{k=1}^{K}\gamma_{m,k}[t]\le D_{\max},\;\forall m,\ \sum_{m=1}^{M}\gamma_{m,k}[t]\le L_{\max},\;\forall k\big\}.
\end{align}
During rollout, the HAP samples a variable-length sequence of feasible UAV--buoy edges and converts it into a binary association matrix. Let $\mathcal T_\gamma[t]$ denote the complete ordered trace, whose selected-edge subsequence is $\big(e^{(1)}[t],\ldots,e^{(Q[t])}[t]\big)$ with $e^{(q)}[t]=(m_q,k_q)\in\mathcal E_{\mathrm{cand}}[t]$. Here, $e^{(q)}[t]$ is the $q$-th selected edge and $Q[t]$ is the number of selected edges. When the policy terminates voluntarily before reaching the capacity bound, $\mathcal T_\gamma[t]$ ends with STOP. The rollout sampler is
\begin{align}\label{eq:assoc_logprob_revised}
&(\mathcal T_\gamma[t],\gamma[t],\log\pi_\Lambda^{\mathrm{seq}}[t])  \\
&=\operatorname{SeqBMatch}_\Lambda\!(
\mathcal E_{\mathrm{cand}}[t],\{\ell_{m,k}[t]\},\ell_{\mathrm{STOP}}[t],D_{\max},L_{\max}),\nonumber
\end{align}
where $\operatorname{SeqBMatch}_\Lambda(\cdot)$ denotes a {sequential stochastic $b$-matching sampler}.
For PPO, $\mathcal T_\gamma[t]$ is the sampled discrete action, whereas $\gamma[t]$ is the deterministic association applied to the environment.
\begin{remark}
The sampler has two useful structural properties. First,
because each selected edge is drawn only from the current
admissible set and all capacity-violating edges are removed
after every update, the resulting binary association always
belongs to \(\mathcal F_\gamma[t]\), including when STOP produces a nonsaturated or empty matching. Second, since the trace is
generated sequentially, its probability factorizes into the product
of the masked-categorical probabilities along the sampled edge and STOP
tokens. Therefore, the association log-probability used by PPO
is exactly the accumulated trace log-probability in
\eqref{eq:assoc_logprob_revised}.
\end{remark}\vspace*{-2mm}

\subsection{Association-to-Motion Directive Realization}\label{subsec:actor}
After the HAP samples the sparse association, the binary matrix \(\gamma[t]\) must be converted into executable local commands. This step is necessary because \(\gamma[t]\) is a global bipartite decision, whereas each UAV only needs the subset of buoys assigned to it, a reference waypoint, and a compact directive context. Moreover, different UAVs may be assigned different numbers of buoys, while the local actor requires a fixed-dimensional observation. We therefore map
each assigned buoy set into a fixed-length slot representation, with dummy slots used when fewer than \(D_{\max}\) buoys are assigned.

\subsubsection{{Directive and Fixed-Length Slot Observation}}
After $\mathrm F_{t-1}$, the HAP selects $\gamma[t]$ and generates $\mathcal D_m[t]=\{\pmb q_m[t],\mathcal C_m[t],\pmb\nu_m^{\mathrm{dir}}[t]\}$, where $\mathcal C_m[t]=\{k:\gamma_{m,k}[t]=1\}$. The compact context $\pmb\nu_m^{\mathrm{dir}}[t]=[\nu_{m,1}^{\mathrm{dir}}[t],\nu_{m,2}^{\mathrm{dir}}[t],\nu_{m,3}^{\mathrm{dir}}[t]]^{\mathrm T}$ is computed only from information available by the end of superframe $t-1$:
\begin{align}\label{eq:nu_revised}
	\nu_{m,1}^{\mathrm{dir}}[t]&={1}/{|\mathcal C_m[t]|}\sum_{k\in\mathcal C_m[t]}\Big[\frac{\Theta_k[t-1]}{\Theta_{\max}}-1\Big]_+,\nonumber \\
	\nu_{m,2}^{\mathrm{dir}}[t]&={|\mathcal C_m[t]|}/{D_{\max}}, \nonumber \\
	\nu_{m,3}^{\mathrm{dir}}[t]&={1}/{|\mathcal C_m[t]|}\sum_{k\in\mathcal C_m[t]}\frac{[\,l_k^{\mathrm{cand}}[t]-L_{\max}\,]_+}{M-L_{\max}}.
\end{align}
Note that these three elements measure the previous realized PCRB excess of the next assigned set, the next UAV-load fraction, and the next candidate-competition pressure, respectively. When $\mathcal C_m[t]=\varnothing$, we set $\pmb\nu_m^{\mathrm{dir}}[t]=\pmb 0$. For the initial directive, $\Theta_k[0]$ is obtained from the common initialized covariance.

The assigned set is then embedded into a fixed-length slot list as $\mathcal I_m[t]=(k_{m,1}[t],\ldots,k_{m,D_{\max}}[t])$ and $\iota_{m,\ell}[t]=\mathbf 1\{k_{m,\ell}[t]\ne 0\}$,
where $k_{m,\ell}[t]\in\{0\}\cup\mathcal C_m[t]$ is the buoy index placed in slot $\ell$, with $k_{m,\ell}[t]=0$ marking an unused dummy slot, so that $\iota_{m,\ell}[t]$ indicates an active slot. Hence, the local actor observation is
\begin{align}\label{eq:obs_revised}
\pmb o_m[t]=\Big[&\pmb c_m[t-1],\pmb q_m[t],\pmb\nu_m^{\mathrm{dir}}[t], \nonumber \\
&\big(\hat{\pmb s}_{k_{m,\ell}}[t|t-1],
\pmb z_{k_{m,\ell}}[t],\iota_{m,\ell}[t]\big)_{\ell=1}^{D_{\max}}\Big].
\end{align}

\subsubsection{{Local Motion and Sensing-Power Actions}}
The local continuous actor is given by
\begin{equation}\label{eq:local_actor_dist_revised}
\pmb a_m^{\mathrm{loc}}[t]
\sim
\pi_{\theta_{\mathrm{loc}}}(\cdot\mid\pmb o_m[t]),
\end{equation}
where $\pmb a_m^{\mathrm{loc}}[t]
=\big[\Delta\pmb q_m^{\mathrm{loc},\mathrm T}[t],\tilde p_{m,1}[t],\ldots,\tilde p_{m,D_{\max}}[t]\big]^{\mathrm T}$, with $\Delta\pmb q_m^{\mathrm{loc}}[t]$ and $\tilde p_{m,\ell}[t]$ being the local waypoint refinement and pre-normalization sensing-power scores of the $D_{\max}$ slots, respectively. The realized constant-altitude position is projected onto the mobility ball, which is given by
\begin{equation}\label{eq:positionupdate_revised}
\pmb c_m[t]
=
\Pi_{\mathcal B(\pmb c_m[t-1],V_{\max}\Delta_{\mathrm T})}
\big(\pmb q_m[t]+\Delta\pmb q_m^{\mathrm{loc}}[t]\big),
\end{equation}
where the projection acts only on the horizontal components and then restores altitude $c_m^z$. The sensing power for active slots is assigned by
\begin{equation}\label{eq:powersplit_revised}
\begin{aligned}
p_{m,k_{m,\ell}}^{\mathrm S}[t]
&=
P_m^{\max}
\frac{\iota_{m,\ell}[t]\exp(\tilde p_{m,\ell}[t])}
{\sum_{j=1}^{D_{\max}}\iota_{m,j}[t]\exp(\tilde p_{m,j}[t])+{\epsilon}},\\
&\ell=1,\ldots,D_{\max},
\end{aligned}
\end{equation}
If $\mathcal C_m[t]=\emptyset$, all sensing powers are set to zero and \eqref{eq:powersplit_revised} is not evaluated. Thus, dummy slots receive no power, and the sensing-power budget and mobility constraint are preserved by construction. The realized PCRB and rate costs are evaluated only after execution and enter the objective through \eqref{eq:reward_revised}.

\subsubsection{{Next-Superframe Directive Generation}}
After $\mathrm F_t$ the HAP constructs $\mathcal G[t+1]$, samples $\gamma[t+1]$, and generates the next directive from the resulting association and one-step predicted buoy states. Define
\begin{equation}\label{eq:wp_buoy_pos_revised}
\hat{\pmb c}_{m,k}^{\mathrm{wp}}[t+1|t]
=\big[[\hat{\pmb s}_k[t+1|t]]_1,[\hat{\pmb s}_k[t+1|t]]_4,c_m^z\big]^{\mathrm T}
\end{equation}
and
\begin{equation}\label{eq:qbar_weight_revised}
\varrho_{m,k}^{\mathrm{wp}}[t+1]
=\lambda_{\Theta}^{\mathrm{sc}}\widehat{\Delta\Theta}_{m,k}[t+1]
+\lambda_R{\widehat R^{\mathrm U}_{m,k}[t+1]},
\end{equation}
where $\lambda_{\Theta}^{\mathrm{sc}},\lambda_R\ge 0$ weight the predicted PCRB-improvement and predicted-rate surrogates. For nonempty $\mathcal C_m[t+1]$, the unprojected waypoint is
\begin{equation}\label{eq:qbar_det_revised}
\tilde{\pmb q}_m[t+1]
=
\frac{\sum_k\gamma_{m,k}[t+1]\varrho_{m,k}^{\mathrm{wp}}[t+1]
\hat{\pmb c}_{m,k}^{\mathrm{wp}}[t+1|t]}
{\sum_k\gamma_{m,k}[t+1]\varrho_{m,k}^{\mathrm{wp}}[t+1]+\epsilon}.
\end{equation}
If the selected set is nonempty but all weights are zero, \eqref{eq:qbar_det_revised} is replaced by the unweighted average over selected buoys. The projected directive waypoint is
\begin{equation}\label{eq:qbar_proj_revised}
\pmb q_m[t+1]
=
\Pi_{\mathcal B(\pmb c_m[t],V_{\max}\Delta_{\mathrm T})}
(\tilde{\pmb q}_m[t+1]),
\end{equation}
while $\pmb q_m[t+1]=\pmb c_m[t]$ if $\mathcal C_m[t+1]=\emptyset$. Finally, the next directive is given by
\begin{equation}\label{eq:nextdirective_revised}
\mathcal D_m[t+1]
=\{\pmb q_m[t+1],\mathcal C_m[t+1],\pmb\nu_m^{\mathrm{dir}}[t+1]\}.
\end{equation}

\subsection{On-Policy Optimization with Centralized Critic}\label{subsec:critic}
The trainable policy parameters are defined as
$\Lambda=\{\psi,\varphi_\ell,\varphi_{\mathrm{STOP}},\theta_{\mathrm{loc}}\}$,
where $\psi$ is the heterogeneous graph encoder, $\varphi_\ell$ and $\varphi_{\mathrm{STOP}}$ are the candidate-edge and STOP logit heads, respectively, and $\theta_{\mathrm{loc}}$ is the local continuous UAV actor. The online optimizer uses exactly the penalized P1 utility $r[t]$ in \eqref{eq:reward_revised}.
%\begin{remark}
%The candidate mask, binary association, UAV load, and
%buoy-cluster constraints are enforced by
%\(\mathcal F_\gamma[t]\) and the masked sequential
%\(b\)-matching sampler. The sensing-power budget is enforced
%by the normalized power split in \eqref{eq:powersplit_revised},
%and the UAV mobility bound is enforced by the projection in
%\eqref{eq:positionupdate_revised}. PCRB, minimum rate, and inter-UAV separation are post-execution soft costs rather than members of the feasible action set. This permits, for example, a transient PCRB excursion to be corrected in later superframes without declaring the sampled action combinatorially infeasible.
%\end{remark}

For centralized training, define the Markov state as
\begin{equation}\label{eq:central_markov_state}
\mathcal S_t^{\mathrm c}=\big(\mathcal G[t],\mathcal X_t^{\mathrm U},
\mathcal X_t^{\mathrm B},\mathcal X_t^{\mathrm S},
\mathcal X_t^{\mathrm{UB}},(t-1)/T,1-(t-1)/T\big),
\end{equation}
where $\mathcal X_t^{\mathrm U}=\{\pmb c_m[t-1],\pmb q_m[t-1],\sum_k\gamma_{m,k}[t-1]\}_{m=1}^{M}$ contains UAV geometry, the previous executed directive, and previous load. The buoy block $\mathcal X_t^{\mathrm B}$ contains $\{\hat{\pmb s}_k[t|t-1],\pmb P_k[t|t-1],\Theta_k[t-1],\widehat\Theta_k^{(0)}[t],b_k[t],a_k[t],u_k\}_{k=1}^{K}$, including each complete belief mean and covariance, realized/prior PCRB summaries, and queue state. The remaining blocks are the sea-patch state $\mathcal X_t^{\mathrm S}=\{\pmb z_s[t]\}_{s=1}^{S}$ and the all-pair context $\mathcal X_t^{\mathrm{UB}}=\{\gamma_{m,k}[t-1],\chi_{m,k}[t],\hat d_{m,k}[t],\widehat R_{m,k}^{\mathrm U}[t],\widehat{\Delta\Theta}_{m,k}[t]\}_{m,k}$. The centralized critic is implemented as a separate value function $V_\phi(\mathcal S_t^{\mathrm c})$. It uses its own graph and permutation-invariant context encoder, and shares no parameters with the association policy.
%The larger state is available only during centralized training. {During execution, the HAP association actor receives $\mathcal G[t]$, whereas each UAV actor receives only $\pmb o_m[t]$ and its directive.}
PPO uses a frozen behavior copy $\Lambda^-$ during rollout, generalized advantage estimation (GAE), clipped actor updates, and value regression. The problem-specific log-probability is written as
\begin{equation}\label{eq:joint_logprob_revised}
\log\pi_\Lambda[t]
:=
\log \pi_\Lambda^{\mathrm{seq}}[t]
+
\sum_{m=1}^{M}\log\pi_{\theta_{\mathrm{loc}}}(\pmb a_m^{\mathrm{loc}}[t]\mid\pmb o_m[t]),
\end{equation}
where $\log \pi_\Lambda^{\mathrm{seq}}[t]$ is the trace log-probability defined in \eqref{eq:assoc_logprob_revised}, including the terminal contribution when the sampled trace ends with STOP.\vspace*{-2mm}

\subsection{Training Workflow}\label{subsec:workflow}
%{The overall architecture and CTDE information flow of the SFPG-MARL framework are summarized in Fig.~\ref{fig:model}. At each superframe, the HAP builds the heterogeneous UAV-buoy coordination graph $\mathcal G[t]$ from current and one-step-predicted quantities, the edge-conditioned encoder $\mathcal F_\psi$ produces the UAV and buoy node embeddings $\{\bm\varpi_m^{\mathrm U}\}$ and $\{\bm\varpi_k^{\mathrm B}\}$, and the edge and STOP heads map the graph to $\{\ell_{m,k}[t]\}$ and $\ell_{\mathrm{STOP}}[t]$. The masked sequential $b$-matching decoder turns these logits into a feasible variable-cardinality association $\bm\gamma[t]$ together with its exact trace log-probability, and the induced directive $\mathcal D_m$ is broadcast to the UAVs, which realize the local motion and sensing-power actions $\bm a_m^{\mathrm{loc}}[t]$ through the shared actor $\pi_{\theta_{\mathrm{loc}}}$. The environment then returns the realized reward $r[t]$, and the centralized complete Markov-state critic $V_\phi(\mathcal S_t^{\mathrm c})$ provides the value baseline to the MAPPO trainer, which updates $\Lambda=\{\psi,\varphi_\ell,\varphi_{\mathrm{STOP}},\theta_{\mathrm{loc}}\}$. The encoder, logit heads, and decoder reside at the HAP and run at both training and execution, whereas the centralized critic is training-only. 
The overall on-policy training is summarized in Algorithm~\ref{alg:psghg_marl}.
Under CTDE, execution needs only the HAP-side association modules and the shared local actor replicated across UAVs. The associated coordination overhead is light, since per superframe each UAV uploads only the local posterior summaries in \eqref{eq:zeta} for its at most $D_{\max}$ served buoys and the HAP broadcasts the compact directive in \eqref{eq:directive_msg}, so the exchanged signaling scales as $\mathcal O(MD_{\max}d_s^2)$ with $d_s$ the buoy state dimension and stays independent of the buoy population beyond the served set.

%Algorithm~\ref{alg:psghg_marl} summarizes the on-policy training workflow. It is included to clarify the information flow of the proposed stochastic association policy. The HAP samples a feasible sparse association, UAVs execute local continuous actions under the received directive, the environment returns the realized sensing, rate, and queue outcomes, and PPO updates the ordered-trace policy using the recorded action log-probability.

\begin{algorithm}[!t]
\caption{On-Policy Training of the Proposed SFPG-MARL}
\label{alg:psghg_marl}
\small
\KwIn{Policy parameters $\Lambda$, critic parameters $\phi$, number of training iterations $N_{\mathrm{it}}$, episodes per iteration $N_{\mathrm{ep}}$, and PPO epochs {$N_{\mathrm{PPO}}$}}
\KwOut{{Learned parameters $\Lambda$ and $\phi$}}
\DontPrintSemicolon
{
\For{$u=1,2,\ldots,N_{\mathrm{it}}$}{
Clear the rollout buffer and freeze the behavior policy $\Lambda^-\leftarrow\Lambda$\;
\For{$n=1,2,\ldots,N_{\mathrm{ep}}$}{
Reset buoy beliefs, UAV states, and traffic states, and set $\pmb q_m[0]=\pmb c_m[0]$\;
Build $\mathcal G[1]$ and $\mathcal S_1^{\mathrm c}$, and sample
$(\mathcal T_{\gamma}[1],\gamma[1])$ under $\Lambda^-$ by
\eqref{eq:assoc_logprob_revised}, and form $\{\mathcal D_m[1]\}$\;
\For{$t=1,2,\ldots,T$}{
Each UAV forms $\pmb o_m[t]$, samples $\pmb a_m^{\mathrm{loc}}[t]$ under
$\Lambda^-$, realizes $\pmb c_m[t]$ and $p_{m,k}^{\mathrm S}[t]$, and
executes $\mathrm S_t$ and $\mathrm U_t$ over $\mathcal C_m[t]$\;
The HAP fuses $\{\zeta_m[t]\}$, evaluates the realized rate and PCRB,
computes $r[t]$, updates the traffic state, and forms the next buoy priors\;
Set $d_t\leftarrow\mathbf 1\{t=T\}$, build $\mathcal G[t+1]$, and form the next
pre-action $\mathcal S_{t+1}^{\mathrm c}$ from the updated causal quantities\;
Store the ordered trace, local actions, behavior log-probability, reward,
causal state transition, terminal flag, and replay context\;
\If{$t<T$}{
Use $\mathcal G[t+1]$ to sample
$(\mathcal T_{\gamma}[t+1],\gamma[t+1])$ under $\Lambda^-$,
and form and broadcast $\{\mathcal D_m[t+1]\}$\;
}
}
}
Set $V_\phi(\mathcal S_{T+1}^{\mathrm c})=0$ and compute GAE advantages
and undiscounted return targets\;
\For{$e=1,2,\ldots,N_{\mathrm{PPO}}$}{
Replay the stored masks, ordered traces, local observations,
and local actions, recompute \eqref{eq:joint_logprob_revised} under $\Lambda$,
update {$\Lambda$} by clipped PPO, and update $\phi$ by value regression\;
}
}}
\end{algorithm}

Furthermore, with $|\mathcal E_{\mathrm{cand}}[t]|$ retained candidate edges and encoder hidden width $d_g$, the graph encoder scales as $\mathcal O(L_g|\mathcal E_{\mathrm{cand}}[t]|d_g^2)$ over $L_g$ message-passing layers and the edge-logit head is linear in $|\mathcal E_{\mathrm{cand}}[t]|$. Let $Q_{\max}[t]\triangleq\min\{MD_{\max},KL_{\max},|\mathcal E_{\mathrm{cand}}[t]|\}$. The sequential decoder uses at most $Q_{\max}[t]+1$ masked-categorical steps, including a possible terminal STOP decision, and costs $\mathcal O((Q_{\max}[t]+1)|\mathcal E_{\mathrm{cand}}[t]|)$. Thus, the online burden scales with the candidate graph rather than the full $MK$ pair space.

\section{Simulations and Results}\label{sec:simulations}
\subsection{{Simulation Setup and Training Configuration}}
We evaluate the proposed framework in the HAP-assisted maritime monitoring scenario described in Section~\ref{sec:systemmodel}. Unless otherwise stated, a $2.5\,\mathrm{km}\times2.5\,\mathrm{km}$ region contains $K=24$ drifting buoys in $S=24$ correlated sea patches and is served by $M=6$ rotary-wing UAVs flying at an altitude of $50\,\mathrm{m}$. The UAVs start from evenly distributed launch positions, and each buoy is initialized in its patch with a small random perturbation. The mission period comprises $T=40$ superframes with $\Delta_{\mathrm T}=1\,\mathrm{s}$. Table~\ref{tab:sim_param_setup} summarizes the default parameters. Later experiments change only the operating condition, network size, or evaluation horizon stated explicitly in the corresponding subsection. For one mission, let $J_q\triangleq\sum_{t=1}^{T}r_Q[t]$ denote the cumulative queue utility, and further define $P_{\Theta}\triangleq\lambda_{\Theta}^{\mathrm{rw}}\sum_{t,k}g_{\Theta,k}[t]$, and $P_R\triangleq\lambda_R^{\mathrm{rw}}\sum_{t,k}g_{R,k}[t]$. Accordingly, the penalized utility is defined as $J_{\mathrm{pen}}\triangleq J_q-P_{\Theta}-P_R$.

Regarding the learning network architectures, the UAV-node, buoy-node, and candidate-edge features are first mapped into a common $64$-dimensional latent space. The HAP-side association actor then applies two rounds of bidirectional edge-conditioned message passing, in which candidate-edge messages are aggregated at their incident UAV and buoy nodes, the node embeddings are updated through residual transformations, and the encoded edge embeddings are formed from the updated endpoint and edge features with Gaussian error linear unit (GELU) activations. A $64$--$64$--$1$ edge head produces one logit for each candidate link, while the STOP head maps the concatenated mean-pooled UAV and buoy embeddings through a $128$--$64$--$1$ multilayer perceptron (MLP). These logits are subsequently processed by the masked sequential $b$-matching decoder. Each UAV employs the same parameter-shared local actor, implemented as a two-hidden-layer MLP with $64$ units per layer and GELU activations, to map its padded local observation and HAP directive into a two-dimensional motion refinement and $D_{\max}$ sensing-power logits. The centralized critic uses an independent two-layer graph encoder together with separate $64$-dimensional encoders for the UAV physical state, buoy belief and PCRB state, sea-patch state, pairwise context, and mission time. The set-valued context blocks are mean- and max-pooled, concatenated with the graph and time representations, and passed through $128$- and $64$-unit fusion layers to produce the scalar state value. The association actor, local actor, and critic do not share parameters. Training uses the Adam optimizer with a learning rate of $3\times10^{-4}$. In addition, we set GAE $\lambda=0.99$, the PPO clipping threshold to $0.2$, and the maximum gradient norm to $1.0$.

For each learning scheme, training uses an aggregate of $43{,}200$ environment superframes across independently initialized runs. Each on-policy iteration collects three $T=40$ episodes and reuses the resulting rollout batch for up to four PPO epochs. During evaluation, the metrics from the independently trained runs are averaged within each unseen test case, and aggregate statistics are then computed across the test cases specified for each experiment.

\begin{table}[!t]
	\centering
	\caption{{Default parameters.}}
	\vspace*{-2mm}
	\label{tab:sim_param_setup}
	\scriptsize
	\renewcommand{\arraystretch}{1.0}
	\setlength{\tabcolsep}{3pt}
	\begin{tabular}{|>{\raggedright\arraybackslash}m{0.27\columnwidth}|>{\raggedright\arraybackslash}m{0.7\columnwidth}|}
		\hline
		\textbf{Category} & \textbf{Parameters and values} \\
		\hline
		Region and mission & Area $2.5\,\mathrm{km}\times2.5\,\mathrm{km}$, $(M,K,S)=(6,24,24)$, $T=40$, $\Delta_{\mathrm T}=1\,\mathrm{s}$ \\
		\hline
		Mobility/cooperation & {$(V_{\max},D_{\max},L_{\max})=(40\,\mathrm{m/s},4,2)$} \\
		\hline
		Sea-patch dynamics & {$(\alpha_q,\beta_q,\sigma_d)=(0.85,0.10,1.5\,\mathrm{km})$, $q\in\{H,\omega,v,\delta\}$} \\
		\hline
		Initial patch state & $H_s[0]\sim\mathcal U[0.5,1.5]\,\mathrm{m}$, $\omega_s[0]\sim\mathcal U[0.4,0.8]\,\mathrm{rad/s}$, $\pmb v_s^{\mathrm{cur}}[0]\sim\mathcal U[-1.5,1.5,0]\,\mathrm{m/s}${, $\delta_s[0]\sim\mathcal U[-0.1,0.1]$} \\
		\hline
		Patch noise & {$(\sigma_{H,s},\sigma_{\omega,s},\sigma_{v_x,s},\sigma_{v_y,s},\sigma_{\delta,s})=(0.05\,\mathrm{m},0.05\,\mathrm{rad/s},0.10\,\mathrm{m/s},0.10\,\mathrm{m/s},0.02)$} \\
		\hline
		Buoy dynamics & $\varsigma_k[t]=\omega_k[t]$, $c_a=0.12$, $\sigma_{a,k}[t]=c_a\omega_k^2[t]H_k[t]$ \\
		\hline
		{Array/carrier} & {$(R,R_2)=(4,16)$, $f_c=5.8\,\mathrm{GHz}$, $\lambda=0.0517\,\mathrm{m}$, $d_{\mathrm a}=\lambda/2$, $B=10\,\mathrm{MHz}$} \\
		\hline
		{Sensing power} & {$P_m^{\max}=33\,\mathrm{dBm}$} \\
		\hline
		Uplink/noise & {$P_k^{\mathrm U}=20\,\mathrm{dBm}$, $\alpha_{\text U}=2.2$, $\beta_0=(\lambda/4\pi)^2$, {$N_0=-104\,\mathrm{dBm}$}} \\
		\hline
		Thresholds & {$(d_{\mathrm{cand}},{\Gamma_{\mathrm{cand}}},\Theta_{\max},R_{\min})=(800\,\mathrm{m},{0.5},10\,\mathrm{m}^2,5\,\mathrm{bit/s/Hz})$} \\
		\hline
		Weights & {$(w_{\mathrm{data}},w_{\mathrm{queue}},w_{\mathrm{pot}})=(0.50,0.25,0.25)$, $(\lambda_{\Theta}^{\mathrm{sc}},\lambda_R)=(0.25,0.20)$,$(\lambda_{\Theta}^{\mathrm{rw}},\lambda_R^{\mathrm{rw}})=(0.1,0.1)$} \\
		\hline
	\end{tabular}
\end{table}

\subsection{Behavior of the Learned Policy}\label{subsec:sim_proposed}
\begin{figure}[!tb]
	\centering
	\includegraphics[width=0.75\columnwidth]{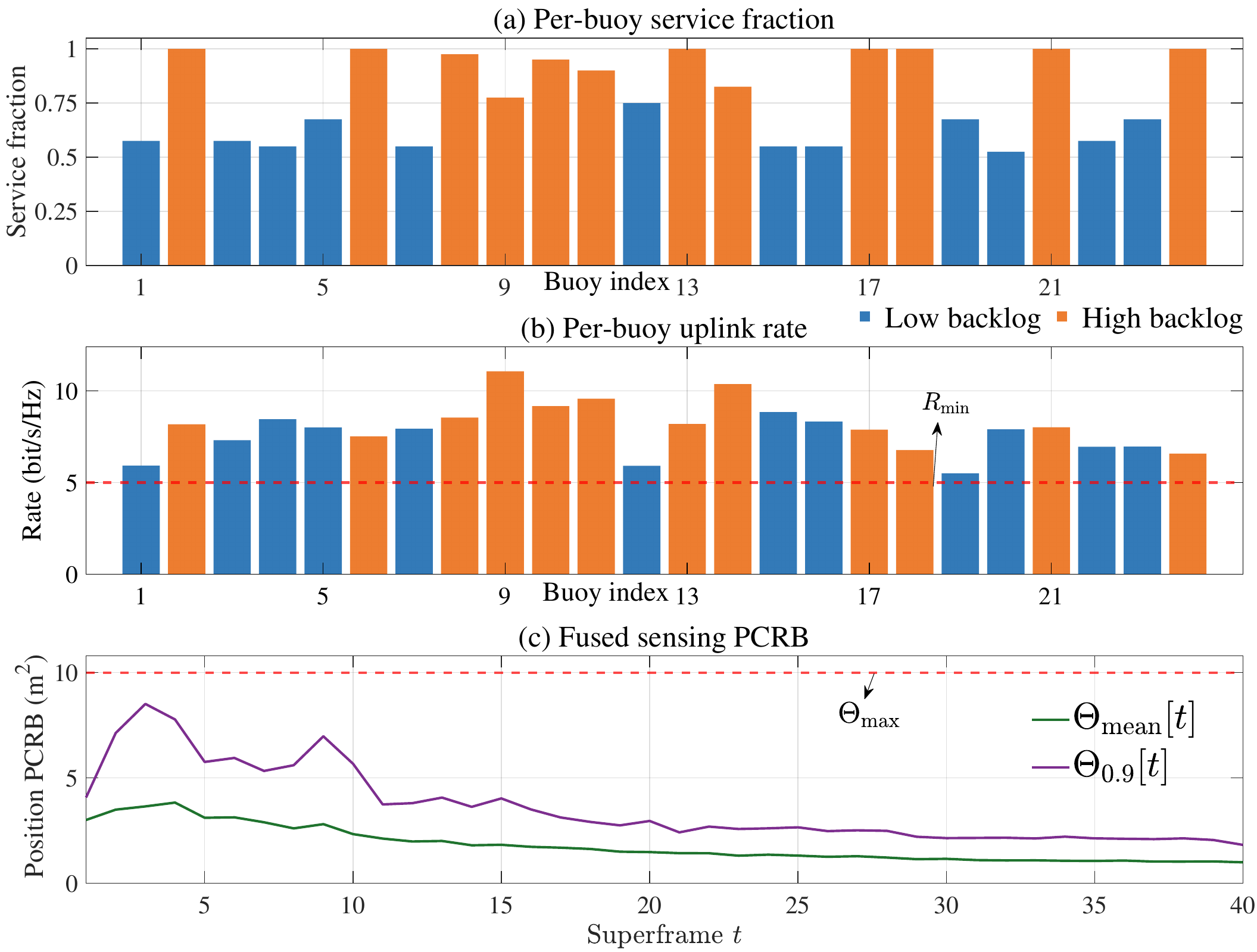}
	\vspace*{-2mm}
	\caption{Per-buoy service and sensing of the converged policy over one deterministic mission with $T=40$ superframes. Orange bars are high-backlog buoys and blue bars are low-backlog buoys. (a)~Per-buoy service fraction. (b)~Per-buoy mean realized uplink rate while served, against the floor $R_{\min}=5$~bit/s/Hz. (c)~Fused position-PCRB, the across-buoy mean $\Theta_\mathrm{mean}[t]$ and the $90$th percentile $\Theta_{0.9}[t]$ against the requirement $\Theta_{\max}=10\,\mathrm m^2$.}
	\label{fig:service_sensing}
\end{figure}

Fig.~\ref{fig:service_sensing} examines how the proposed policy distributes service across buoys while maintaining communication and sensing quality. For buoy $k$, the service fraction in Fig.~\ref{fig:service_sensing}(a) is $f_k=T^{-1}\sum_t\mathbf 1\{\sum_m\gamma_{m,k}[t]\geq1\}$. On average, the high-backlog group is served about $1.6$ times as often as the low-backlog group, while every buoy is served in more than half of the superframes. The conditional mean realized rate in Fig.~\ref{fig:service_sensing}(b) remains above $R_{\min}$ for every buoy. Fig.~\ref{fig:service_sensing}(c) further shows that the across-buoy 90th-percentile PCRB stays below $\Theta_{\max}$ and decreases markedly by the end of the mission. Together, these results show that the learned policy prioritizes urgent queues without sacrificing service coverage or sensing quality.

\begin{figure}[!tb]
	\centering
	\includegraphics[width=0.7\columnwidth]{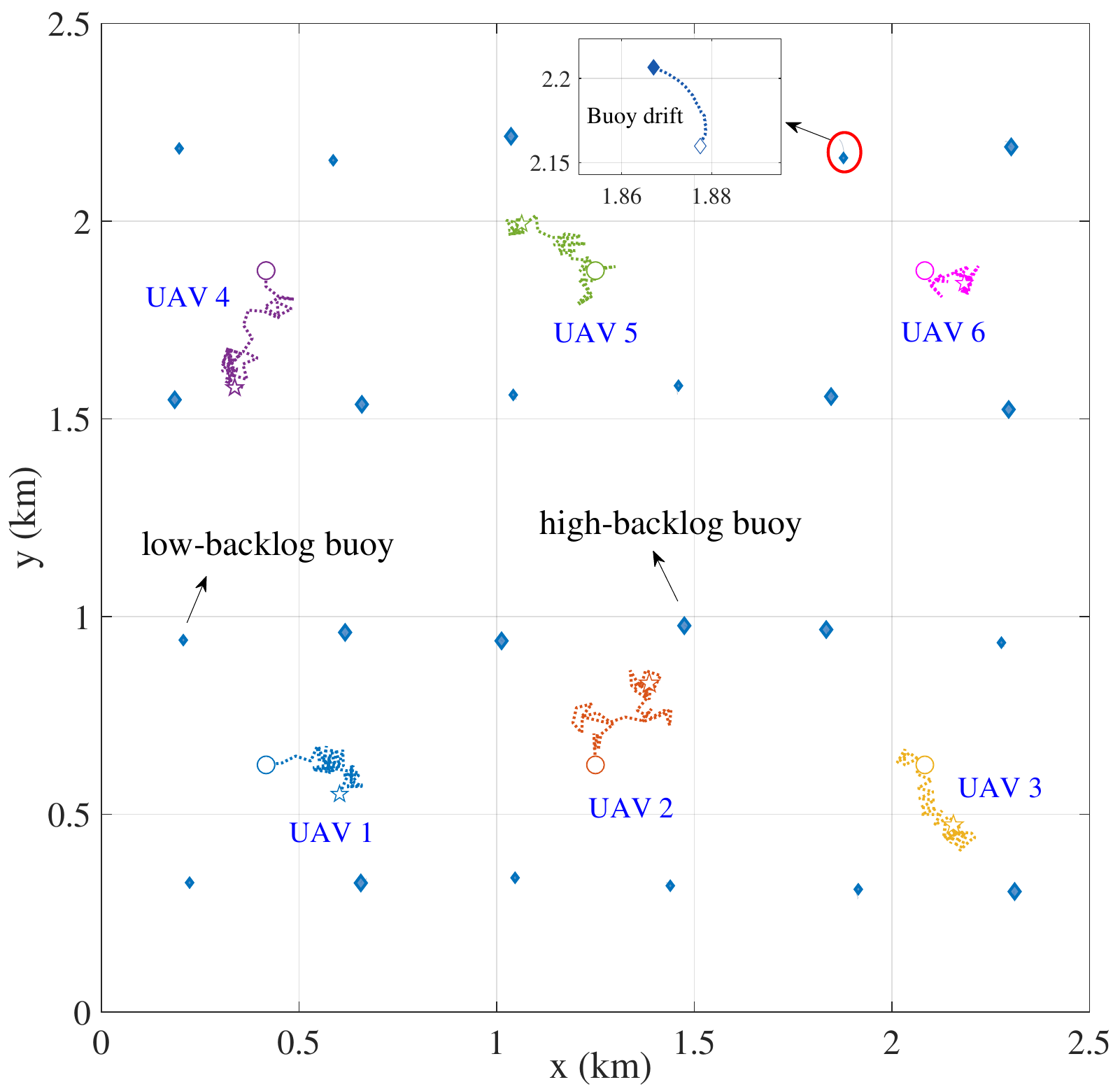}
	\vspace*{-2mm}
	\caption{{Long-horizon UAV trajectories of the converged policy. Each UAV moves from its launch point ($\circ$) to its final position ($\star$). The blue diamond markers denote buoys and are sized by their initial backlogs. The inset shows the buoy with the largest displacement.}}
	\label{fig:mission_motion}
\end{figure}

{Fig.~\ref{fig:mission_motion} further illustrates the dynamic UAV--buoy coordination through the long-horizon UAV trajectories. The UAVs first reposition toward clusters of heavily backlogged buoys and then refine their positions as the buoys drift. The inset confirms that buoy displacement accumulates over the mission, while Fig.~\ref{fig:service_sensing} quantifies the corresponding service and sensing behavior.}

\begin{table}[!t]
	\centering
	\caption{{Comparison over unseen cases under the default setting.}}
	\vspace*{-2mm}
	\label{tab:main_comparison}
	\scriptsize
	\renewcommand{\arraystretch}{1.08}
	\setlength{\tabcolsep}{2.2pt}
	\begin{tabular}{l c c c c}
		\hline
		Method & $J_q$ & $P_{\Theta}$ & $P_R$ &  $J_{\mathrm{pen}}$ \\
		\hline
		RAND & $4.15$ & $6.70$ & $0.58$ &  $-3.14\pm2.77$ \\
		HOVER & $4.49$ & $3.42$ & $0.00$ &  $1.08\pm9.65$ \\
		MW & $9.79$ & $8.01$ & $0.10$ &  $1.67\pm1.02$ \\
		CA-MW & $9.66$ & $3.70$ & $0.07$ &  $5.89\pm0.74$ \\
		MQO & $10.07$ & $6.32$ & $0.01$ &  $3.75\pm0.69$ \\
		E-MLP & $6.36$ & $6.37$ & $0.17$ &  $-0.19\pm6.43$ \\
		{DQN} & {$9.41$} & {$1.54$} & {$0.03$} &  $7.84\pm0.89$ \\
		{GNN-FS} & $8.59$ & $3.21$ & $0.10$ &  $5.28\pm0.47$ \\
		{DTDE} & $8.79$ & {$1.81$} & $0.02$ &  $6.96\pm0.68$ \\
		Matched-SAC & $8.50$ & $1.15$ & $0.08$ & $7.27\pm1.84$ \\
		GNN-Greedy & $9.87$ & $3.16$ & $0.06$ &  $6.66\pm0.51$ \\
		\textbf{Proposed} & $\mathbf{10.47}$ & $\mathbf{0.02}$ & $0.07$ &  $\mathbf{10.38\pm0.77}$ \\
		\hline
	\end{tabular}
\end{table}
\begin{table}[!t]
	\centering
	\caption{{Cross-size generalization from $(M,K)=(6,24)$ to $(8,32)$, $(10,40)$, and $(12,48)$.}}
	\vspace*{-2mm}
	\label{tab:generalization}
	\scriptsize
	\renewcommand{\arraystretch}{1.08}
	\setlength{\tabcolsep}{2.5pt}
	\begin{tabular}{l c c c}
		\hline
		\multirow{2}{*}{Method} & \multicolumn{3}{c}{$J_{\mathrm{pen}}$} \\
		\cline{2-4}
		& $(8,32)$ & $(10,40)$ & $(12,48)$ \\
		\hline
		E-MLP & $-6.89\pm13.44$ & $-2.60\pm1.74$ & $-22.74\pm23.37$ \\
		DQN & $7.16\pm1.60$ & $9.80\pm0.99$ & $8.97\pm1.63$ \\
		GNN-FS & $4.93\pm0.76$ & $5.93\pm0.80$ & $3.45\pm1.29$ \\
		DTDE & $4.32\pm0.92$ & $0.50\pm0.39$ & $-5.66\pm5.69$ \\
		Matched-SAC & $9.07\pm2.16$ & $14.16\pm0.80$ & $14.86\pm2.22$ \\
		GNN-Greedy & $7.02\pm1.54$ & $9.64\pm0.31$ & $8.19\pm1.56$ \\
		\textbf{Proposed} & $\mathbf{11.19\pm2.70}$ & $\mathbf{17.54\pm0.82}$ & $\mathbf{17.85\pm1.77}$ \\
		\hline
	\end{tabular}
\end{table}

\begin{table}[!t]
	\centering
	\caption{{Longer-mission generalization at $T=100, 150, 200$.}}
	\vspace*{-2mm}
	\label{tab:horizon_robustness}
	\scriptsize
	\renewcommand{\arraystretch}{1.08}
	\setlength{\tabcolsep}{2.5pt}
	\begin{tabular}{l c c c}
		\hline
		\multirow{2}{*}{Method} & \multicolumn{3}{c}{$J_{\mathrm{pen}}$} \\
		\cline{2-4}
		& $T=100$ & $T=150$ & $T=200$ \\
		\hline
		E-MLP & $2.55\pm8.49$ & $2.89\pm10.78$ & $3.75\pm10.80$ \\
		DQN & $19.86\pm5.18$ & $29.41\pm10.58$ & $36.99\pm16.42$ \\
		GNN-FS & $15.35\pm2.49$ & $22.05\pm4.30$ & $28.11\pm6.01$ \\
		DTDE & $15.66\pm1.83$ & $20.77\pm3.49$ & $23.72\pm6.03$ \\
		Matched-SAC & $14.27\pm3.22$ & $18.58\pm4.28$ & $21.28\pm6.63$ \\
		GNN-Greedy & $21.42\pm4.92$ & $33.15\pm10.26$ & $44.39\pm16.35$ \\
		\textbf{Proposed} & $\mathbf{26.31\pm5.24}$ & $\mathbf{39.57\pm8.61}$ & $\mathbf{51.61\pm11.96}$ \\
		\hline
	\end{tabular}
\end{table}

\begin{table*}[!b]
	\centering
	\caption{{Performance under significant-wave-height and sustained-current variations.}}
	\label{tab:operating_sweeps}
	\scriptsize
	\renewcommand{\arraystretch}{1.05}
	\setlength{\tabcolsep}{2.4pt}
	\begin{tabular}{c c c c c c c c c c c}
		\hline
		\multirow{2}{*}{Factor} & \multirow{2}{*}{Level}
			& \multicolumn{3}{c}{CA-MW} & \multicolumn{3}{c}{DQN} & \multicolumn{3}{c}{Proposed} \\
		\cline{3-11}
		& & $J_{\mathrm{pen}}$ & $P_{\Theta}$ & $P_R$
		& $J_{\mathrm{pen}}$ & $P_{\Theta}$ & $P_R$
		& $J_{\mathrm{pen}}$ & $P_{\Theta}$ & $P_R$ \\
		\hline
		\multirow{3}{*}{\shortstack{mean $H_s[0]$\\(m)}}
			& $1.0$ & $4.60\!\pm\!0.99$ & $4.98$ & $0.07$ & $6.42\!\pm\!1.06$ & $2.96$ & $0.03$ & $\boldsymbol{8.80\!\pm\!0.87}$ & $1.60$ & $0.07$ \\
			& $1.5$ & $2.64\!\pm\!1.02$ & $7.04$ & $0.11$ & $4.53\!\pm\!1.02$ & $4.99$ & $0.03$ & $\boldsymbol{6.54\!\pm\!0.91}$ & $3.85$ & $0.06$ \\
			& $2.0$ & $0.29\!\pm\!1.20$ & $9.37$ & $0.09$ & $2.18\!\pm\!1.07$ & $7.34$ & $0.03$ & $\boldsymbol{4.32\!\pm\!0.93}$ & $6.14$ & $0.07$ \\
		\hline
		\multirow{3}{*}{\shortstack{sustained current\\mean (m/s)}}
			& $2$ & $4.88\!\pm\!1.09$ & $4.93$ & $0.11$ & $6.43\!\pm\!1.08$ & $2.95$ & $0.04$ & $\boldsymbol{8.71\!\pm\!1.01}$ & $1.60$ & $0.06$ \\
			& $3$ & $4.83\!\pm\!1.02$ & $4.97$ & $0.13$ & $6.41\!\pm\!1.04$ & $3.01$ & $0.06$ & $\boldsymbol{8.70\!\pm\!0.95}$ & $1.62$ & $0.10$ \\
			& $4$ & $4.63\!\pm\!0.83$ & $5.01$ & $0.18$ & $6.13\!\pm\!0.87$ & $3.03$ & $0.09$ & $\boldsymbol{8.59\!\pm\!0.70}$ & $1.61$ & $0.12$ \\
		\hline
	\end{tabular}
\end{table*}

\subsection{Comprehensive Comparison with Baselines}\label{subsec:sim_baselines}

\subsubsection{{Benchmark Schemes}}
{We compare SFPG-MARL with randomized and deterministic schedulers, learning-based alternatives, and two structural ablations. The benchmark schemes are summarized below.}
\begin{itemize}[leftmargin=1.5em,itemsep=0pt,topsep=2pt,parsep=0pt]
	\item \textbf{RAND}. Random priorities are assigned to the candidate UAV--buoy edges before feasible matching. This scheme provides a state-independent association reference under the same structural constraints.
	\item \textbf{HOVER}. Each UAV hovers at the launch position, while exact association follows the predicted rates. It provides a fixed-geometry mobility reference.
	\item \textbf{MW}~\cite{Tassiulas1992MaxWeight}. Exact maximum-weight matching combines queue backlogs with predicted service rates, but does not account for sensing quality.
	\item \textbf{CA-MW}. A causal estimate of the marginal PCRB improvement is added to MW, yielding the primary sensing-aware deterministic benchmark.
	\item \textbf{MQO}. The instantaneous penalized utility is maximized in each superframe. This benchmark represents myopic optimization without explicitly accounting for future queue and geometry evolution.
	\item {\textbf{E-MLP}}~\cite{Shen2023GNNWireless}. {A shared edge-wise MLP replaces graph message passing, while the sequential association rule and STOP action are retained. It evaluates coordination without relational feature aggregation.}
	\item{\textbf{DQN}}~\cite{Hu2025MADQNLAE}. {A centralized DQN assigns action values to candidate edges, after which feasible matching produces the association. It represents centralized value-based allocation under the same training-interaction budget.}
	\item \textbf{GNN-Greedy}. The learned graph representation and local policy of SFPG-MARL are retained, but the prefix-conditioned variable-cardinality decoder is replaced by greedy edge selection. This comparison highlights the role of structured association decoding.
	\item \textbf{GNN-FS}~\cite{Wang2022GNNResource}. Following the use of GNNs for wireless resource allocation, a heterogeneous graph encoder is trained and combined with fixed-cardinality feasible selection.
	\item {\textbf{DTDE}}~\cite{deWitt2020IPPO}. {Each UAV follows a recurrent policy learned from local observations, while capacity conflicts are resolved by deterministic arbitration. This scheme represents independent local learning without explicit graph-based coordination.}
	\item \textbf{Matched-SAC}~\cite{Haarnoja2018SAC}. It retains the graph representation, prefix-conditioned association decoder, action space, and interaction budget of SFPG-MARL, while replacing the PPO updates with the maximum-entropy off-policy SAC update.
\end{itemize}
All independently trained learning schemes follow the common budget in Section~V-A, with DQN using replay-based updates, DTDE using recurrent PPO updates, and Matched-SAC differing from SFPG-MARL only in the online policy-update rule.

\subsubsection{{Training Convergence}}
\begin{figure}[!tb]
	\centering
	\includegraphics[width=0.75\columnwidth]{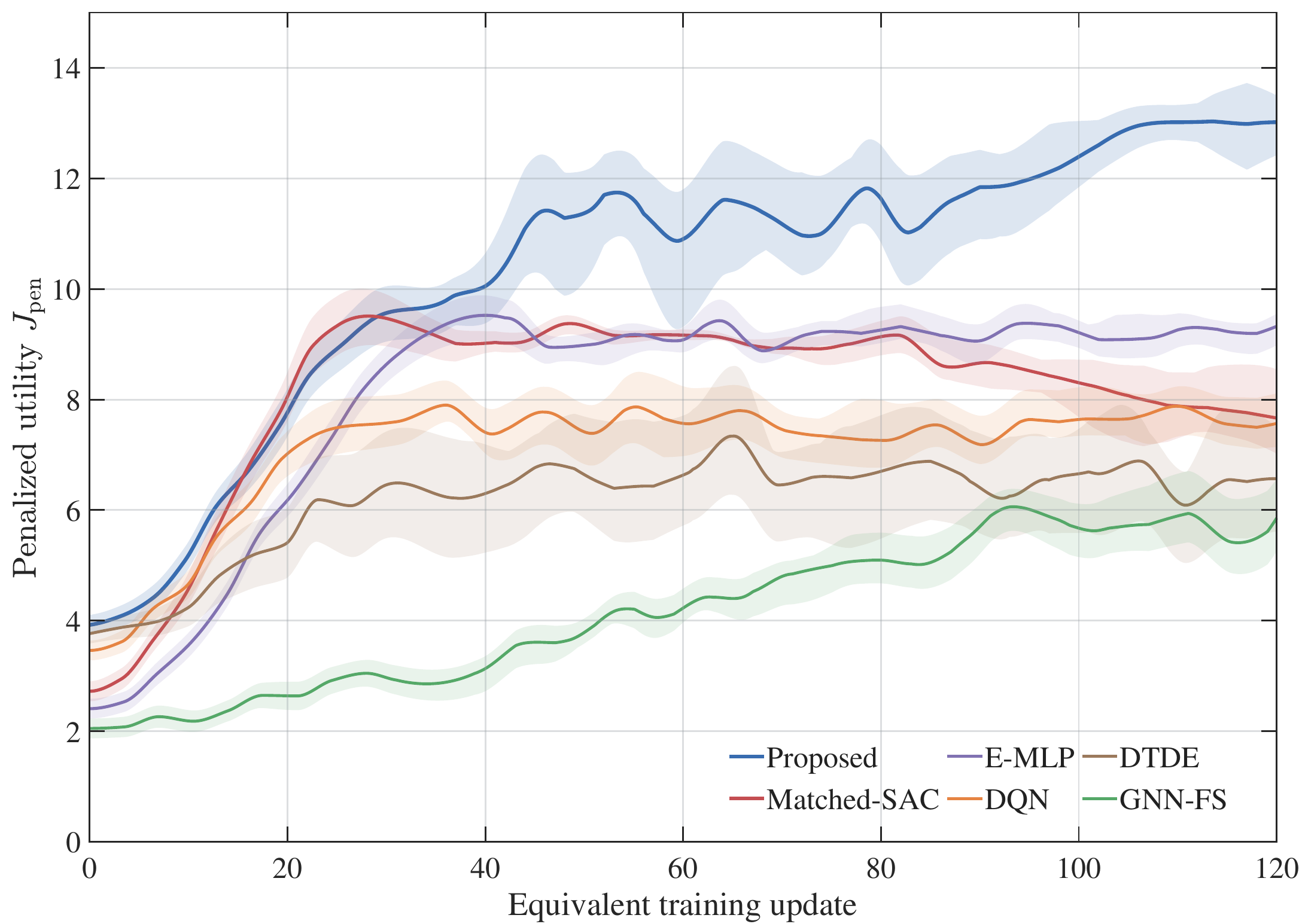}
	\vspace*{-2mm}
	\caption{Training curves of the learning designs.}
	\label{fig:convlearn}
\end{figure}
Fig.~\ref{fig:convlearn} shows rapid early improvement for most learning-based methods, followed by slower adaptation near their respective plateaus. SFPG-MARL fluctuates during the intermediate stage but resumes improving and reaches the highest performance. E-MLP stabilizes at an intermediate level, DQN and DTDE converge lower, Matched-SAC gives back part of its early gain, and GNN-FS improves more gradually.

\subsubsection{{Performance Comparison}}
As shown in Table~\ref{tab:main_comparison}, SFPG-MARL attains a penalized utility of $10.38$, about $32\%$ higher than DQN and $76\%$ higher than CA-MW. It combines the largest queue utility of $10.47$ with a weighted PCRB penalty of only $0.02$, whereas the high queue utility of MQO is offset by a much larger sensing penalty. The ablations localize the contributions of the policy construction. GNN-Greedy retains the learned graph representation and local policy but replaces prefix-conditioned adaptive decoding with greedy selection, reducing $J_{\mathrm{pen}}$ to $6.66$. E-MLP retains the sequential decoder and STOP action but removes graph message passing and yields a negative mean, while fixed-cardinality GNN-FS reaches $5.28$. Matched-SAC changes only the update rule and reaches $7.27$. Together, these comparisons support complementary benefits from relational aggregation, variable-cardinality structured association, and PPO updates.

\subsection{Robustness and Generalization}

\subsubsection{{Generalization to a Larger Network}}
{Table~\ref{tab:generalization} directly applies the policies learned at $(M,K,T)=(6,24,40)$ to the larger $(8,32,40)$, $(10,40,40)$, and $(12,48,40)$ networks without further training. All enlarged settings retain the same area and physical parameters, while progressively increasing the numbers of UAVs, buoys, and feasible association choices.}

As shown in Table~\ref{tab:generalization}, SFPG-MARL remains the highest-performing method at all three enlarged network sizes. In the largest network with $12$ UAVs and $48$ buoys, its penalized utility reaches $17.85$, about $20\%$ higher than Matched-SAC and nearly twice that of DQN. GNN-Greedy remains positive, whereas DTDE and E-MLP become negative at the largest scale and GNN-FS remains substantially lower. The consistent ranking across progressively denser topologies demonstrates reliable zero-shot transfer of the structured graph policy.

\subsubsection{{Generalization to a Longer Mission}}
The policies learned with $T=40$ are next applied directly to missions of length $T=100$, $T=150$, and $T=200$ without additional training. The network size and physical parameters remain unchanged, while the longer durations require the policies to manage queue and geometry evolution over progressively more superframes.

As shown in Table~\ref{tab:horizon_robustness}, SFPG-MARL retains the highest cumulative penalized utility at every longer horizon. {Over the longest mission of $200$ superframes, its penalized utility reaches $51.61$, compared with $44.39$ for GNN-Greedy and $36.99$ for DQN, while E-MLP exhibits much larger variability.} The learned policy therefore preserves its advantage as the mission extends well beyond the training horizon.

\subsubsection{Robustness to Maritime Conditions}
Table~\ref{tab:operating_sweeps} evaluates three significant-wave-height profiles and three direction-preserving sustained-current levels. The wave-height sweep spans moderate conditions and, in the highest profile, local rough-sea realizations under the Douglas scale~\cite{WMO2026Douglas}. The current sweep increases the mean speed while retaining the native directions and relative spatial variation.

Increasing wave height reduces the penalized utility of all three methods, primarily through a rising PCRB penalty. This trend follows the model because the buoy-dynamics uncertainty is scaled by $\sigma_{a,k}[t]=c_a\omega_k^2[t]H_k[t]$ following \eqref{eq:buoydyn}, while the sea-state-indexed clutter term in \eqref{eq:localsense} increases the effective echo noise in \eqref{eq:local_fim} and hence the fused PCRB in \eqref{eq:pcrb}. SFPG-MARL retains the highest utility and the lowest PCRB penalty at every level. At the largest initial mean significant wave height of $2.0\,\mathrm{m}$, its penalized utility remains $4.32$, compared with $2.18$ for DQN and $0.29$ for CA-MW.

The sustained-current sweep produces much smaller changes. Under direction-preserving current scaling, the current velocity enlarges the translational drift term in \eqref{eq:buoydyn}, whereas the normalization in \eqref{eq:patchtransport} leaves the directional transport factor nearly unchanged. The sweep therefore changes buoy displacement without substantially altering the cross-patch transport pattern. As the mean current increases from $2$ to $4\,\mathrm{m/s}$, the penalized utility of SFPG-MARL decreases by only $1.4\%$, and its PCRB penalty remains near $1.60$. The method therefore retains the best sensing--service balance throughout this range.

\section{Conclusion}\label{sec:conclusion}
This paper formulated HAP-assisted maritime sparse cooperative ISAC as a finite-horizon queue-weighted collection problem that couples buoy drift, sensing accuracy, sparse cooperation, UAV mobility, and buffered data service. SFPG-MARL combines heterogeneous graph coordination with prefix-conditioned variable-cardinality association and parameter-shared local motion and sensing-power control. Under the default setting, it provides the best sensing--service balance, with clear gains over both learning-based and deterministic alternatives and a nearly negligible PCRB penalty. Ablations confirm complementary contributions from relational aggregation, adaptive association, and PPO updates. The same policy transfers without retraining to larger networks and longer missions and remains robust across the considered wave-height and sustained-current ranges. These results demonstrate effective coordination across changing topology, mission duration, and maritime dynamics. Future work will consider imperfect HAP--UAV control links and field-calibrated maritime channel and motion models.

\raggedbottom

\bibliographystyle{IEEEtran}
\bibliography{ocean_drl_gnn}

\end{document}